\documentclass[aps,prc,reprint,noshowpacs,groupedaddress,onecolumn]{revtex4}
\usepackage{bm}
\usepackage{amsmath}
\usepackage{amsfonts}
\usepackage{amscd}
\usepackage{graphicx}
\usepackage{color}
\def\beq{\begin{equation}}
\def\eeq{\end{equation}}

\begin{document}

\title{Scattering using real-time path integrals}

\author{W.~N.~Polyzou}
\affiliation{Department of Physics and Astronomy, The University of
Iowa, Iowa City, IA 52242}

\author{Ekaterina Nathanson}
\affiliation{School of Science and Technology, Georgia Gwinnett
College, Lawerenceville, GA 30042}

\date{\today}


\begin{abstract}

  {\bf Background:} Path integrals are a powerful tool for solving
  problems in quantum theory that are not amenable to a treatment by
  perturbation theory.  Most path integral computations require an
  analytic continuation to imaginary time.  While imaginary time
  treatments of scattering are possible, imaginary time is not a
  natural framework for treating scattering problems.  More importantly,
  quantum algorithms for calculating path integrals require real-time
  evolution.

  {\bf Purpose:} To test a recently introduced method for performing direct
  calculations of scattering observables using real-time path
  integrals in order to understand the challenges facing real-time
  path integral calculations of scattering observables.  

  {\bf Method:} The computations are based on a new interpretation of
  the path integral as the expectation value of a potential functional
  on cylinder sets of continuous paths with respect to a complex probability
  distribution.  The method can in principle be applied
  to arbitrary short-range potentials.

  {\bf Results:} The method is applied to compute
  matrix elements of M{\o}ller wave operators applied to narrow wave
  packets.  These are used to calculate half-shell sharp-momentum
  transition matrix elements for one-dimensional potential scattering.
  The calculations for half-shell transition operator matrix elements
  converge to the numerical solution of the Lippmann-Schwinger equation.  

  {\bf Conclusions:} This work presents a proof in principle that
  scattering observables can be computed using real-time
  Feynman path integrals.  While the computational method is not efficient,
  it can be improved.  It provides a laboratory for studying quantum
  computational algorithms that are applicable to scattering problems.

\end{abstract}

\maketitle

\section{Introduction}

The purpose of this paper is to explore the possibility of using
real-time path integral methods \cite{Feynman_1}\cite{Feynman_2} to
calculate scattering observables.  The proposed computational method
is based on a recent formulation of the path integral
\cite{Muldowney}\cite{Katya_1}\cite{Katya_2} that replaces the
integral over paths by the expectation value of the potential
functional, $F[\gamma] := e^{-i \int V(\gamma(t)) dt}$, with respect
to a complex probability distribution on a space of paths.  The space
of paths is cylinder sets of paths with fixed starting and end points.
The method is not limited to quadratic potentials; it can be used to
compute real-time path integrals for arbitrary short-range potentials.

The motivation for this work is that path integrals provide one of the
most reliable methods to study strongly interacting quantum
systems. In cases of physical interest the path integral is
analytically continued to imaginary time, so the integrals can be
approximated using Monte-Carlo methods \cite{Ulam}\cite{Metropolis}.
The most important experimental observables are scattering
observables.  While clever methods have been developed to extract
scattering observables from Euclidean path integrals
\cite{Luscher}\cite{Luscher2}\cite{Lee}\cite{Luu}\cite{Gordon},
imaginary time is not the most natural setting for computing
scattering observables.

In the future it is anticipated that path integral computations will
be performed by quantum computers.  These computations require real-time
evolution.  While the technology for quantum computation is not
mature, an investigation of real-time methods for the computation
of path integrals is warranted.  These investigations can be used to
study the efficiency of real-time computational methods in
anticipation of future developments in quantum computing.  While most
applications of real-time path integrals to scattering are limited to
generating perturbative expansions \cite{Faddeev}, there are related
works on the formulation of scattering theory based on real-time path
integrals \cite{Rosenfelder1} \cite{Rosenfelder2} \cite{Campbell}.

This paper investigates the application of the formulation of the path
integral in \cite{Muldowney}\cite{Katya_1}\cite{Katya_2} to directly
calculate scattering observables in real time.  While this test is
limited to potential scattering in one dimension, the computation uses
strongly convergent time-dependent methods that are applicable to
general scattering reactions in quantum mechanics and quantum field
theory.  While the calculations presented in this paper are not
applicable to these complex reactions, the long term goal of quantum
information science is be able to treat realistic reactions, so it
make sense to investigate methods that are not limited to the one-dimensional
case.

Compared to Euclidean methods, which use Monte Carlo integration, the
real-time path integral in this work is approximated by matrix elements of
powers of a large matrix, which can be computed efficiently.
The matrix approximates a unitary transfer matrix that generates the
contribution due to multiple paths in parallel, one time step at a time.



The theory of complex probabilities was first contemplated by R.
Henstock \cite{Henstock}\cite{bartle}, based on a generalized theory
of integration that he co-developed with J. Kurzweil.  The
Henstock-Kurzweil integral is similar to a Riemann integral, except
the usual $\epsilon$-$\delta$ definition is replaced by: for any
prescribed error, $\epsilon$, there is a positive function,
$\delta (x)$ (called a gauge), where the intervals $I_n$ and
evaluation points $x_n$ in the Riemann sums satisfy
$I_n \in [x_n -\delta(x_n) , x_n+\delta (x_n)]$. In the
Henstock-Kurzweil case the points and intervals are correlated or
tagged and the evaluation point is normally taken as one of the
endpoints of the interval.  Henstock-Kurzweil integral reduces to the
Riemann integral when the gauge function, $\delta (x)$, is a constant.
 
One advantage of the Henstock-Kurzweil integral is that it can still
be approximated to arbitrary accuracy by generalized Riemann sums.
The class of Henstock-integrable functions contains the Lebesgue
measurable functions as well as functions that are not absolutely
integrable.  These considerations generalize to infinite dimensional
integrals, which are inductive limits of finite dimensional integrals
over cylinder sets that have a well-defined volume and correlated
evaluation points and times.  Precise definitions for the interested
reader can be found in \cite{Muldowney}\cite{Katya_1}\cite{Katya_2}.


In classical probability theory the probability is a non-negative
function on a collection of sets.  It has the property that the
probability that a measurement will fall in a set or its complement is
1 and the probability that a measurements will fall in a countable
union of disjoint sets is the sum of the probabilities that the
measurement will fall in each set.  If the sets are generated by
intervals under complements and countable unions, the collection of
sets are Lebesgue measurable sets and probability is a Lebesgue
integrable function.  Because Henstock-Kurzweil integrable functions
include Lebesgue integrable functions, Henstock proposed the
replacement of a probability theory based on countably additive
positive measures by one based on {\it finitely} additive generalized
Riemann sums.  The two formulations of probability coincide for real
probabilities, but the Henstock formulation of probability extends to
non-positive and complex probabilities.  This was further explored by
P. Muldowney.  He developed a suggestion of Henstock that the Feynman
path integral could be understood using this framework.  Muldowney
\cite{Muldowney} proved that the method converges to a local solution
of the Schr\"odinger equation.  It was recently shown by
P. J{\o}rgensen and one of the authors \cite{Katya_1}\cite{Katya_2}
that the local solutions could be patched together to construct a
global solution of the Schr\"odinger equation, and a unitary
one-parameter time-evolution group.  The unitary one-parameter
time-evolution group is used to formulate the scattering problem.  In
addition, the support of the complex probability is on paths that are
almost everywhere continuous \cite{Katya_2}, leading to a rigorous
interpretation of the real-time path integral as the expectation of a functional
of continuous paths with respect to a complex probability
distribution.

The approach taken in this paper is a numerical implementation of the
formulation of the path integral as the expectation of a potential
functional with respect to a complex probability distribution.  Gill
and Zachary \cite{Gill} investigated an alternative direct application of the
Henstock-Kurzweil integral to path integrals.  

While this is just a reinterpretation of the ordinary path integral,
it leads to a computational method that is not limited to quadratic
potentials.  In this paper this method is applied to treat a
one-dimensional scattering problem with a Gaussian potential.  While
this is a simple problem, from a path integral perspective it involves
approximating a large-dimensional oscillating integral.  The
computational method used in this work computes the transition matrix
elements in a straightforward, but inefficient way.  The efficiency of
the method can be significantly improved, although it is unlikely to be
competitive with standard methods using ordinary computers.

An appealing property of this method is that the interpretation of the
calculation as the expectation value of a potential functional with
respect to a complex probability distribution on a space of continuous
paths is not lost in the numerical implementation.  In particular the
computational algorithm can be deconstructed to find the approximate
complex probability for any given cylinder set of paths.  

Since the discussion that follows is somewhat detailed, a brief
summary that outlines the essential elements of the computational
method is given below.  The new method starts like the ordinary path
integral.  It is based on the Trotter product formula \cite{Simon},
which is used to express the unitary time-evolution operator as a
strong limit
\beq
e^{-iHt} = \lim_{N \to \infty} (e^{-i {p^2\over2\mu}{t \over N} }e^{-iV{t \over N} })^N
\label{i.1}
\eeq
provided that ${{p}^2\over 2 \mu} +V$ is essentially self-adjoint.  It
has recently been shown that the limit exists for
a class of Hamiltonians with singular 
potentials that have self-adjoint extensions \cite{Katya_3}.

The next step is to insert complete sets of intermediate states that
separately diagonalize both the kinetic energy, $p^2/2\mu$, and
potential energy, $V$.  The integrals over the intermediate momenta
are Gaussian Fresnel integrals which can be performed analytically.
What remains is the large $N$ limit of $N$-dimensional integrals over
the real line. Each integral is interpreted as an integral over all
space at a given time slice.  These steps are standard \cite{Feynman_2}
and can be
found in any textbook that covers path integrals.  The new steps are:
\begin{itemize}

\item[1.)] Replace each of the $N$ integrals over the real line by a sum
  of integrals over small intervals.  These intervals represent
  windows that a path passes through at each time slice.
  The limit that the width of the finite intervals vanish, and the finite
  endpoints of the semi-infinite intervals become infinite is
  eventually taken.  For computational purposes the intervals should
  be sufficiently small so the potential is approximately
  constant on each interval and is approximately zero on the semi-infinite
  intervals.

\item[2.)] The next step is to perform the integrals over the different
  products of N intervals, one for each time slice, {\it assuming that
    the potential is zero}.  The results are complex quantities that
  are labeled by $N$ space intervals, $\{I_{n_1}, \dots I_{n_N}\}$;
  one interval for each time slice.  Each sequence of intervals
  defines a cylinder set.  A cylinder set is an ordered set of $N$
  windows at $N$ intermediate times.  Every path goes
  through a unique cylinder set.  Thus each cylinder set represents an
  equivalence class of paths.  It is elementary to show that the
  sum of these integrals over all possible N-fold products of
  intervals (cylinder sets) is 1.  This allows these integrals to be
  interpreted as complex probabilities that a path is an element of
  the associated cylinder set.

\item[3.)] If the intervals are sufficiently small so the potential is
  approximately locally constant on each interval, then the sum of
  products of the complex probabilities with
  $e^{-i \sum V(x_i) \Delta t}$, where $x_i$ is any sample point in
  the $i^{th}$ interval in the cylinder set, converges to the ``path
  integral'' defined by the Trotter product formula.  Because the potential is
  only needed at a finite set of sample points, this last step makes
  the method applicable to any-short range potential,

\end{itemize}

The new interpretation is that the ill-defined real-time path integral is
replaced by a well-defined expectation value of the potential
functional with respect to a complex probability distribution on
cylinder sets of continuous paths.  This brief summary is discussed in
more detail in what follows.

There are two problems that must be overcome to make this into a
computational method.  They can be summarized by noting that (1) it is
not clear that the complex probabilities can be computed analytically;
they involve $N$-dimensional integrals and (2) even if they could be
computed either analytically or numerically, there are too many of
them.  If there are $M$ intervals in each of $N$ time slices, the number of
cylinder sets is $M^N$, where the final result is obtained in the
limit that both $M$ and $N$ become infinite.

The challenge of this work is to overcome these obstacles
without giving up the interpretation of the path integral as the
expectation value of a functional of paths with respect to a complex
probability distribution of paths.

The virtue of this formulation of the ``path integral'' is that the
non-existent path integral is replaced by the well-defined expectation
value of a potential functional of continuous paths with respect to a
complex probability distribution on cylinder sets of continuous paths.

While the test problem treated in this paper can be computed more
efficiently by directly solving the Lippmann-Schwinger equation, path
integrals are a powerful tool for solving problems in quantum field
theory that are not limited by perturbation theory.  While the methods
discussed in this paper do not directly apply to the field theoretic
case,
they provide a laboratory for testing computational strategies that
could be implemented in the future on a quantum computer.

The paper is organized as follows.  The next section includes a brief
discussion of the scattering formalism that will be used in the rest
of the paper.  This formalism is based on standard time-dependent
scattering theory. The time-dependent method has the advantage that it
is in principle applicable to multi-particle scattering in three
dimensions, scattering with long-range potentials \cite{Dollard}
\cite{dollard:1978we} and scattering in quantum field theory
\cite{Haag:1958vt}\cite{Ruelle:1962}.  Using a method that can be
applied to a large class of scattering reactions is preferred for
developing algorithms that might be implemented on a quantum computer
in the future.  Section three discusses the Feynman path integral
formulation of the scattering problem.  The fourth section introduces
the reinterpretation of the path integral as the expectation value of
the potential functional with respect to a complex probability on a
space of continuous paths.  Section five introduces a factorization of
the complex probabilities that makes numerical computations possible.
Section six analyzes the test calculation for scattering from a simple
short-range potential in one dimension.  A summary and concluding
remarks appear in section seven.
 
\section{Scattering observables using time-dependent methods}

The application that will be considered in this work is scattering in
one-dimension using an attractive Gaussian potential, $V(x) = -v_0e^{-(x/r_0)^2}$.
The goal is to calculate sharp-momentum transition matrix elements
using a path-integral treatment of time-dependent scattering.  The
sharp-momentum transition matrix elements are simply related to the
scattering cross section.  The method can be formally generalized to
compute sharp-momentum transition matrix elements for multi-particle
scattering in three-dimensions.  Because this method is ultimately
based on the Trotter product formula, which requires a strong limit,
the desired matrix elements need to be extracted using narrow wave
packets.

In quantum mechanics the probability for scattering from an initial
state $\vert \psi_i \rangle$ to a final state $\vert \psi_f \rangle$,
is
\beq
P = \vert \langle \psi_f (t) \vert \psi_i (t) \rangle \vert^2 =
\vert \langle \psi_f (0) \vert \psi_i (0) \rangle \vert^2 .
\label{s.1}
\eeq
The time-independence of the scattering probability follows from the
unitarity of the time-evolution operator.  Since this probability is
independent of time, it can be computed at any convenient common time.
In a scattering experiment the initial state, $\vert \psi_i (t)
\rangle$, is simple long before the collision; it looks like a system
of free moving particles, $\vert \psi_{i0} (t) \rangle$.  Similarly,
the final state, $\vert \psi_f (t) \rangle$, has a simple form long
after the collision; it looks like a system of free moving particles,
$\vert \psi_{f0} (t) \rangle$.  The difficulty is that there is no
common time when both states have a simple form.  Time-dependent
scattering theory provides a means to express the initial and final
scattering states at a common time in terms of states of
asymptotically free particles at a common time.  The free particle
states are easily computed at any time.

The relation of the initial and final states to the asymptotic system
of free particles, described by $\vert \psi_{f0}(t) \rangle$ and
$\vert \psi_{i0}(t) \rangle$, is given by the scattering asymptotic
conditions:
\beq
\lim_{t \to \infty} \Vert \vert \psi_f(t) \rangle - \vert \psi_{f0}(t)
\rangle \Vert =0 
\qquad
\lim_{t \to -\infty} \Vert \vert \psi_i(t) \rangle - \vert \psi_{i0}(t)
\rangle \Vert =0 .
\label{s.2}
\eeq
In the general multichannel or field theoretic case these expressions
have the form
\beq
\lim_{t \to \infty} \Vert e^{-iHt} \vert \psi_f(0) \rangle - 
\Phi_\alpha e^{-iH_\alpha t}\vert \psi_{f\alpha }(0)
\rangle \Vert =0 
\qquad
\lim_{t \to -\infty} \Vert e^{-iH t} \vert \psi_i(0) \rangle - 
\Phi_\beta e^{-iH_\beta t}\vert \psi_{i\beta}(0)
\rangle \Vert =0 
\label{s.3a}
\eeq
where $H$ is the Hamiltonian of the system, $\alpha$ and $\beta$
represent the final and initial scattering channels, $\Phi_{\alpha}$
and $\Phi_{\beta}$ are injection operators that map spaces of
asymptotically free point particles into the physical Hilbert space by
including the internal structure of the asymptotic particles,
$H_\alpha$ and $H_\beta$ are the energies of the free asymptotic
particles and $\vert \psi_{i\beta}(0)\rangle$ and
$\vert \psi_{i\beta}(0)\rangle $ are products of sharp momentum wave
packets in the total momentum and spin of each asymptotic fragment.
In quantum field theories the $\Phi_{\alpha/ \beta}$ are products of
the quasilocal functions of fields that create one-body states out of
the vacuum in Haag-Ruelle theory \cite{Haag:1958vt}\cite{Ruelle:1962}.
There are also $\Phi_{\alpha/ \beta}$'s for Coulomb and other
long-range interactions \cite{Dollard} \cite{dollard:1978we}.

The unitarity of $e^{iHt}$ can be
used to write this as
\beq
\lim_{t \to \infty} \Vert\vert \psi_f(0) \rangle - 
e^{iHt}\Phi_\alpha e^{-iH_\alpha t}\vert \psi_{f\alpha }(0)
\rangle \Vert =0 
\qquad
\lim_{t \to -\infty} \Vert \vert \psi_i(0) \rangle - 
e^{iHt} \Phi_\beta e^{-iH_\beta t}\vert \psi_{i\beta}(0)
\rangle \Vert =0 .
\label{s.3b}
\eeq
The important point is that in all cases the limits are strong and the
dynamics enters via $e^{-iHt}$ which can in principle be computed by a
path integral.  The only non-trivial input is the channel injection
operators $\Phi_\alpha$, which are products of bound states or
products of interpolating fields that create point-mass eigenstates
out of the vacuum.

For two-body scattering (\ref{s.3b})  
has the form
\beq
\lim_{t \to \infty} \Vert \vert \psi_f(0) \rangle - 
e^{iHt} e^{-iH_0t} \vert \psi_{f0}(0)
\rangle \Vert =0 
\qquad
\lim_{t \to -\infty} \Vert \vert \psi_i(0) \rangle - 
e^{iHt} e^{-iH_0t}  \vert \psi_{i0}(0)
\rangle \Vert =0 
\label{s.4}
\eeq
where $H=H_0+V$ is the Hamiltonian of the two-body system.
These formulas express the initial and final scattering states at the
common time, $t=0$, in terms of the corresponding non-interacting states at
the same time.

While these expressions formally involve the limits $t\to \pm \infty$,
if $t=0$ is taken as the time of the collision, the limit is realized
at the finite times when $\pm t$ are large enough for the interacting
particles to be outside of the range of the interaction.  In a real
experiment the times when the beam and target are prepared and when
the reaction products are seen in detectors are finite; the infinite-time
limits are a simple way to ensure that $t$ is large enough to reach
the limiting form.  This means that in a realistic calculation the
limit can be replaced by a direct evaluation at a sufficiently large
finite $t$ or $-t$.  The minimum size of this $t$ depends on the range
of the interaction, the size of the wave packets, and momentum
distribution in the wave packet.  When finite times are used in
calculations, the minimal size of $t$ needs to be determined for each
calculation.

Because it appears in equation (\ref{s.4}) it is useful to define the
operator
\beq
\Omega (t) := e^{iHt} e^{-iH_0t} .
\label{s.5}
\eeq
The operator $\Omega (t)$ is a unitary operator that, for sufficiently
large $\pm t$, transforms the initial or final {\it non-interacting} wave
packet at time $t=0$ to the initial or final interacting wave packet at
$t=0$.  Equation (\ref{s.4}) shows
\beq
\vert \psi_i(0) \rangle \approx \Omega (-t) \vert \psi_{i0}(0) \rangle 
\label{s.5b}
\eeq
as $t$ gets sufficiently large.

The quantities of physical interest are the on-shell
transition matrix elements.   For the simplest case of
scattering by a short-range potential, $V(x)$, the transition matrix elements
are related to $\Omega (t)$ by 
\beq
\langle \mathbf{p}_f \vert T(E_i+i0^+) \vert \mathbf{p}_i \rangle
=
\lim_{t \to - \infty}
\langle \mathbf{p}_f \vert V \Omega (t) \vert \mathbf{p}_i \rangle
\label{s.6}
\eeq
where
\beq
T(z) = V +V (z-H)^{-1}V
\label{s.7}
\eeq
is the transition operator and
$E_i = {\mathbf{p}_i^2 \over 2m}= {\mathbf{p}_f^2 \over 2m}$.
Because the limit on the right is a strong limit, it only exists if 
the sharp-momentum generalized eigenstate, $\vert \mathbf{p}_i \rangle$, is
replaced by normalizable a wave packet.

A useful approximation is to use a narrow Gaussian wave packet, centered 
about the initial momentum, $\mathbf{p}_i$, with a delta-function normalization:
\beq
\int d\mathbf{p} \langle \mathbf{p} \vert \psi_{i0}\rangle =1 .
\label{s.8}
\eeq
With this choice \cite{Lippmann}, 
\beq
\langle \mathbf{p}_f \vert T(E_i) \vert \mathbf{p}_i \rangle
 \approx
\langle \mathbf{p}_f \vert V \Omega (-t) \vert \mathbf{p}_i \rangle
\approx
\langle \mathbf{p}_f \vert V \Omega (-t) \vert \psi_{i0}(0) \rangle
\label{s.9}
\eeq
in the large $t$, narrow wave packet limit.  Gaussian wave packets are
minimal uncertainty states, which provide maximal control over both
the momentum resolution and spatial width of the free wave packet.
For computational purposes, the width of the wave packet in momentum
should be chosen so
$\langle \mathbf{p}_f \vert T(E_i) \vert \mathbf{p}_i \rangle$ varies
slowly on the support of the wave packet.

In this work ``path integral'' methods are used to compute the
right-hand side of equation (\ref{s.9}).  Because the path integral is
formulated in terms of paths in the coordinate representation, the
actual quantity that needs to be computed (in one dimension) is the Fourier
transform
\beq
\langle {p}_f \vert V \Omega (-t) \vert \psi_{i0}(0) \rangle =
{1 \over \sqrt{2 \pi}}
\int dx e^{-ip_fx} \langle {x} \vert V \Omega (-t) \vert \psi_{i0}(0) \rangle
\label{s.10}
\eeq
for a sufficiently large $t$ and narrow wave packet.  This can be
computed by a direct computation of the Fourier transform or by
integrating over a narrow final-state wave packet,
$\langle \psi_{f0}(0) \vert$, centered about ${p}_f$ with a
delta-function normalization:
\beq
\langle {p}_f \vert V \Omega (-t) \vert \psi_{i0}(0) \rangle \approx 
\langle \psi_{f0}(0) \vert V \Omega (-t) \vert \psi_{i0}(0) \rangle .
\label{s.11}
\eeq

The transition matrix elements in (\ref{s.9}) are directly related to 
scattering matrix elements,  cross sections,
phase shifts and, in this one-dimensional case,
to reflection and transmission coefficients.  In the 
one-dimensional case scattering matrix elements
are related to transition matrix elements by
\beq
\langle p \vert S \vert p' \rangle =
\delta (p-p') - 2 \pi i \delta (E-E') 
\langle p \vert T(E+i0^+) \vert p' \rangle  
\label{s.12}
\eeq
where $p=\pm p'$.
The phase shifts are defined by
\beq
\langle E(\pm) \vert S \vert E'\rangle = \delta (E-E') e^{2 i \delta_{\pm} (E)}
= \sqrt{{m \over p}} \langle p \vert S \vert p' \rangle  \sqrt{{m \over p'}}
.
\label{s.13}
\eeq
There are two values 
depending on whether the final state has momentum parallel or anti-parallel
to the initial state.

Using $\vert p \rangle = \vert E \rangle \sqrt{{p\over m}}$ in
(\ref{s.12}) gives
\beq
e^{2 i \delta_+ (E)}= 1 - 2 \pi i {m \over \vert p \vert }  
\langle p \vert T(E+i0^+) \vert p \rangle
\label{s.14}
\eeq
\beq
e^{2 i \delta_- (E)}= - 2 \pi i {m \over \vert p \vert }  
\langle -p \vert T(E+i0^+) \vert p \rangle .
\label{s.15}
\eeq
This leads to the following expressions for the phase shifts in terms of 
the real and imaginary parts of the on-shell transition matrix elements:
\beq
\delta_{+} (E) = {1 \over 2} \tan^{-1} ({-2 \pi m Re(\langle  p \vert
T(E+i0^+) \vert p
\rangle) 
\over \vert p\vert  + 2 \pi m 
Im (\langle  p \vert T(E+i0^+) \vert p \rangle)})
\label{s.16}
\eeq
\beq
\delta_{-} (E) = {1 \over 2} \tan^{-1} ({- Re(\langle - p \vert T (E+i0^+)
\vert p
\rangle) 
\over  
Im (\langle - p \vert T(E+i0^+) \vert p \rangle)}) .
\label{s.16}
\eeq
The on-shell transition matrix elements are related to the transmission
and reflection coefficients by
\beq
T=1-{2 \pi i p \over m} \langle p \vert T(E+i0^+) \vert p \rangle
\qquad
R= -{2 \pi i p \over m} \langle -p \vert T(E+i0^+) \vert p \rangle .
\label{s.16}
\eeq
The expression $\vert R\vert^2 + \vert T \vert^2=1$ expresses the unitarity
of the scattering operator
\beq
S_{++}^* S_{++} + S_{+-}^* S_{-+} =1
\label{s.16}
\eeq

\section{ Scattering using the Feynman path integral} 

The dynamical quantity needed as input to equation (\ref{s.9}) is 
\beq
\langle x \vert V \Omega (-t) \vert \psi_{i0}(0) \rangle =
\langle x \vert V
e^{-iHt} \vert
\psi_{i0} (-t) \rangle .
\label{fs.1}
\eeq
for sufficiently large $t$.  The initial state at time zero is a 
Gaussian approximation to a delta function with the initial momentum,
$p_i$:
\beq
\langle p \vert \psi_{i0}  (0) \rangle =
{1 \over 2 \sqrt{\pi} \Delta p} e^{- {(p-p_i)^2 \over 4 (\Delta p)^2}} .
\label{fs.2}
\eeq
Here $\Delta p$ is the quantum mechanical uncertainty in $p$ for this
wave packet.  This wave packet needs to be evolved to $-t$ using the 
free time evolution.  The result is
\beq
\langle p \vert \psi_{i0} (-t) \rangle =
{1 \over 2 \sqrt{\pi} \Delta p} e^{ 
- {(p-p_i)^2 \over 4 (\Delta p)^2} + i {p^2 \over 2 \mu}t}.
\label{fs.3}
\eeq
These wave packets are needed in the coordinate basis in the path 
integral.  The Fourier transform of (\ref{fs.3}) can be computed 
analytically and expressed in terms of the momentum or coordinate uncertainty
of the wave packet:
\beq
\langle x \vert \psi_{i0} (t) \rangle =
(2 \pi)^{-1/2}
\sqrt{{1  \over 1  + i {2 (\Delta p)^2 t \over  \mu}}}
e^{-
{(\Delta p)^2 \over 1  + {4 (\Delta p)^4 t^2 \over  \mu^2}}
(x-{p_it\over \mu})^2}
e^{i{1 \over 1  + {4 (\Delta p)^4 t^2 \over  \mu^2}}
(x p_i   
+ 2 x^2 (\Delta p)^4 {t \over  \mu}
- { t \over  \mu}{p_i^2 \over 2 })} =
\label{fs.4}
\eeq
\beq
(2 \pi)^{-1/2}
\sqrt{{1  \over 1  + i { t \over 2 (\Delta x)^2  \mu}}}
e^{-{1 \over 4 \Delta x^2}
{1 \over 1  + { t^2 \over 4 \Delta x^4   \mu^2}}
(x-{p_it\over \mu})^2}
e^{i{1 \over 1  + {t^2 \over 4 \Delta x^4  \mu^2}}
(x p_i   
+ { x^2 t \over  8 \Delta x^4 \mu}
- { t \over  \mu}{p_i^2 \over 2 })} .
\label{fs.5}
\eeq
where $\Delta p \Delta x = {1 \over 2}$, since Gaussian wave functions 
represent minimal uncertainty states.  Equations (\ref{fs.4})
and (\ref{fs.5}) have the
form a Gaussian with a moving center multiplied by an oscillating
function.
These equations show that the center of this
initial wave packet moves with its classical velocity, $v=p_i/\mu$, so
the center of the wave packet is located at
${x}(t) = { {p}_i \over \mu}t$, where $\mu$ is the mass and $p_i$ is
the mean momentum of the initial wave packet.  Ignoring the spreading
of the wave packet, it will be in the range of the interaction for a
time $t \approx {(R + \Delta x) \mu \over \vert \mathbf{p}_i \vert}$,
where $R$ is the range of the potential and $\Delta x$ is the width of
the wave packet.  This suggests that the asymptotic time for
scattering will be reached for
$t > {(R+ \Delta x)\mu \over \vert {p}_i\vert} $.  Because the
potential appears in (\ref{s.9}) in the expression for the transition
matrix elements, only the values of
$\langle {x} \vert e^{-iHt} \vert \psi_i(-t) \rangle$ inside the range
of the potential are needed to calculate the transition matrix
elements.


Equation (\ref{fs.1}) can be expressed exactly as
\beq
\langle x \vert
V e^{-iHt} \vert 
\psi_{i0} (-t) \rangle =
\lim_{N \to \infty} 
\langle x \vert V
\left (e^{-iH{t/N}}\right )^N  \vert
\psi_{i0} (-t) \rangle . 
\label{fs.6}
\eeq
In the ``$x$'' representation
the wave function in (\ref{fs.6}) has the form
\beq
\int \langle x \vert e^{-i ({p^2 \over 2\mu} + V) t} \vert x_i \rangle
dx_i \langle x_i \vert \psi_{i0} (-t)\rangle  = 
\lim_{N\to \infty}
\int \langle x \vert (e^{-i ({p^2 \over 2\mu} + V)\Delta t})^N \vert x_i
\rangle dx_i \langle x_i \vert \psi_{i0} (-t)\rangle . 
\label{fs.7}
\eeq
where $\Delta t := t/N$. 

The only contributions to the large $N$ limit come from the
first-order terms in $\Delta t$.  This follows from the Trotter
product formula \cite{Simon}, which gives conditions for the operator
version of
\beq
e^x = \lim_{N \to \infty}(1+{x/N})^N 
\label{fs.8}
\eeq
to hold when $N\to \infty$ as a strong limit.

Using this property, the limit in (\ref{fs.7}) can be replaced by
\beq
\lim_{N\to \infty}
\int \langle x \vert 
\underbrace{e^{-i {p^2 \over 2\mu}\Delta t}
e^{-i V \Delta t}
\cdots 
e^{-i {p^2 \over 2\mu}\Delta t}
e^{-i V \Delta t}}_{N-\mbox{times}}
\vert x_i 
\rangle   dx_i \langle x_i \vert \psi_{i0} (-t)\rangle .
\label{fs.9}
\eeq
The following replacements were used in (\ref{fs.7}) 
to get (\ref{fs.9})
\beq
e^{-i ({p^2 \over 2\mu} + V)\Delta t} \to
1 -i ({p^2 \over 2\mu} + V)\Delta t  \to
(1 -i {p^2 \over 2\mu}\Delta t )(1-i V \Delta t )\to
e^{-i {p^2 \over 2\mu}\Delta t}
e^{-i V\Delta t} .
\label{fs.10}
\eeq
This representation has the advantage that unitary time evolution
is expressed as the limit of products of unitary operators.

The next step is to insert complete sets of intermediate position 
and momentum eigenstates
so ${p^2 \over 2\mu}$ and $V$ each become multiplication operators.  This 
leads to the expression 
\beq
\langle x_0 \vert
e^{-iHt} \vert
\psi_{i0} (-t) \rangle =
\lim_{N\to \infty}
\int \prod_{n=1}^N {dp_n dx_n \over 2\pi} 
e^{ i p_i (x_{n-1}-x_n) - i {p_n^2 \over 2\mu}\Delta t -i 
V(x_n) \Delta t} \langle x_N \vert \psi_{i0} (-t)\rangle .
\label{fs.11}
\eeq

The $p_n$ integrals are Gaussian Fresnel integrals and can be performed by
completing the square in the exponent
\beq
\langle x_0 \vert
e^{-iHt} \vert
\psi_{i0} (-t) \rangle =
\lim_{N\to \infty}
\int \prod_{n=1}^N {dp_n dx_n \over 2 \pi}
e^{
-i {\Delta t \over 2\mu} (p_n - {\mu \over \Delta t}(x_{n-1}-x_n))^2
+i {\mu \over 2\Delta t} (x_{n-1}-x_n)^2
-i 
V(x_n) \Delta t }
\langle x_N \vert \psi_{i0} (-t)\rangle. 
\label{fs.12}
\eeq
The general structure of the resulting momentum integrals is 
\beq
\int_{-\infty}^{\infty} e^{-i a p^2 + i bp } dp =
\sqrt{\pi \over i a} e^{i b^2/(4a) }.
\label{fs.13}
\eeq
These integrals are computed by completing the square in the exponent,
shifting the origin, and evaluating the resulting integral by contour
integration over a pie shaped path with one edge along the positive
real $p$ axis and the other making a $45$ degree angle between the positive 
real and negative imaginary $p$ axis.

The resulting integral over the $N$ momentum variables is 
\[
\langle x_0 \vert e^{-iHt}  \vert \psi_{i0} (-t) \rangle =  
\]
\beq
\lim_{N\to \infty}
({\mu \over 2 \pi i \Delta t})^{N/2}
\int \prod_{i=n}^N dx_n
e^{
i {\mu \over 2\Delta t} (x_{n-1}-x_n)^2
-i V(x_n) \Delta t }
\langle x_N \vert \psi_{i0} (-t)\rangle.  
\label{fs.14}
\eeq
This is the standard form of the path integral derived by Feynman. 
The path integral interpretation is obtained by factoring 
a $\Delta t$ out of the sum in the exponent to get
\beq
\langle x_0 \vert e^{-itH}  \vert \psi_{i0} (-t) \rangle =  
\lim_{N\to \infty}
({\mu \over 2 \pi i \Delta t})^{N/2}
\int 
e^{
i \sum_{n=1}^N \left ({\mu \over 2 } ({x_{n-1}-x_n\over \Delta t})^2
- V(x_n)\right ) \Delta t }\prod_{m=1}^N dx_m
\langle x_N \vert \psi_{i0} (-t)\rangle . 
\label{fs.15}
\eeq
This looks like an integral over piece-wise linear paths between
points in the $N$ time slices, $(x_N \to x_{N-1} \to \cdots \to x_0)$,
weighted with the imaginary exponential of a finite difference 
``approximation'' of the action:
\beq
A \approx  \sum_{n=1}^N \left ({\mu \over 2 } ({x_{n-1}-x_n\over \Delta t})^2
- V(x_n)\right ) \Delta t .
\label{fs.16}
\eeq
This is in quotes because, due to the integrals, the numerator in the
finite difference does not get small as $\Delta t \to 0$, so the
interpretation of ${x_{n-1}-x_n\over \Delta t}$
as an approximate derivative is not justified.

Irrespective of any concerns about the interpretation, this expression
is mathematically well-defined as a limit of finite dimensional
integrals.  It converges as a result of the Trotter product formula,
however it is not very useful for computational purposes because of
the large dimensionality of the integrals needed for convergence,
except in the case of quadratic interactions, where the integrals can
be computed analytically.

\section{The Muldowney-Nathanson-J{\o}rgensen path integral}

To compute the path integral for scattering it is necessary to
overcome several obstacles.  These include the large dimensionality of
the integrals, the need to compute with general short-range
interactions, the oscillatory nature of the integrals,
and the spreading of the scattering wave packets.  The purpose of
this work is to investigate some methods that have the potential to
overcome these obstacles.

The proposed computations are a consequence of the reformulation of
the path integral due to Muldowney \cite{Muldowney}, Nathanson and 
J{\o}rgensen \cite{Katya_1}\cite{Katya_2}.  This
provides a means for treating a large class of interactions, and
eliminates the questionable ``finite difference'' approximation of the
kinetic energy in (\ref{fs.15}).  There still remain oscillations
associated with the potential term; but they are only relevant inside
of the finite range of the potential.

This method replaces the usual interpretation of the path integral by
assigning a ``complex probability'' to subsets of 
paths, and computing
the expectation value of the random variable $F[\gamma] = e^{-i \int
V(\gamma (t) ) dt}$ with respect to this complex probability
distribution.  In this expression $\gamma(t)$ is a path
between $x_N$ and $x_0$ , This differs from the standard
interpretation in that the action functional is replaced by a
potential functional, and the ``measure'' is replaced by a complex
probability distribution.  The random variable $e^{-i \int V(\gamma
  (t) ) dt}$ differs from 1 only on the portion of the path, $\gamma
(t)$, that is in the range of the potential.  The potential functional
does not suffer from the interpretational difficulties of the action
functional in the standard path integral.

This is still a computationally intractable problem.  In order to make
this computationally tractable, the complex probability is
approximately factored into a product of complex probabilities for
each ``time step''. These one-step ``complex probabilities'' have the
advantage that they can be computed analytically and that the analytic
calculation treats the free propagation exactly.  The important
simplification is that the one-step complex probability can be
approximated by a matrix and the multi-step probability is
approximately the $N$-fold product of the same matrix.  This reduces
the calculation of the transition matrix elements to the computation
of powers of a matrix applied to a vector.  In this case the usual
Monte Carlo integration is replaced by matrix multiplication, which
can be performed efficiently. 

Finally, the use of the operator $\Omega (t)$ in (\ref{s.5}) means
that the quantity being computed in (\ref{s.6}) and (\ref{s.9}) is a
deformation of the initial free wave packet, at the time of collision,
to the corresponding interacting packet at the same time.  In this
case both the free and interacting wave packets remain localized near
the origin and the parameter $t$ interpolates between the free and
interacting localized states.  The spreading of the wave packet is
only relevant during the time of collision, and even during that time
some cancellations are expected.  Thus,
the relevant parts of the calculation take place in a finite
space-time volume.

The fundamental new idea that is the key to the computational strategy,
proposed by Muldowney, is to decompose the integral over each
$x_n$ in (\ref{fs.14}) into a sum of integrals over $M_n+1$ intervals,
$I_{mn}$,
at the $n^{th}$ time slice:
\beq
\int dx_n = \sum_{m=0}^{M_n} \int_{I_{mn}} dx_n .
\label{njm.1}
\eeq
The intervals $I_{mn}$ are chosen to be disjoint and cover the real line.  They
are taken to have the  general form
\beq
\underbrace{(-\infty, x_{1n})}_{I_{0n}}, 
\underbrace{[x_{1n},x_{2n})}_{I_{1n}} , \cdots , 
\underbrace{[x_{M-1,n},x_{M,n})}_{I_{M-1,n}},
\underbrace{[x_{M,n},\infty)}_{I_M,n} .
\label{njm.2}
\eeq
Using this decomposition the limit in (\ref{fs.15}) becomes
\beq
\langle x_0 \vert e^{-itH}  \vert \psi_{i0} (-t) \rangle =  
\lim_{N\to \infty} 
({\mu \over 2 \pi i \Delta t})^{N/2}
\sum_{m_1 \cdots m_N}
\prod_{n=1}^N  \int_{I_{m_n}} dx_n
e^{
i {\mu \over 2\Delta t} (x_{n-1}-x_n)^2
-i V(x_n) \Delta t }
\langle x_N \vert \psi_{i0} (-t)\rangle.  
\label{njm.3}
\eeq
The sum is over the $(M_1+1)\times \cdots \times (M_N+1)$ $N$-fold
Cartesian products of intervals (cylinder sets) for the $N$ time
slices.  Each continuous path from $x_N$ to $x_0$ goes through one
interval in each time slice and is thus an element of a unique
cylinder set.  The $m_n$ sums range over $0\leq m_n \leq M_n$, $1 \leq
n \leq N$.  Up to this point everything is independent of how the
intervals are chosen.   For potentials and initial wave packets that are
smooth, it is enough to choose the
intervals sufficiently small so that the interaction and initial free
wave packet are approximately constant on each interval, $I_{mn}$.
Then the contribution from the potential can be factored out of the integral
over the interval, and be replaced by evaluating the potential at any
point $y_{mn} \in I_{mn}$ in the interval.  In the calculations exhibited in
section 6, $y_{mn}$ is taken to be the midpoint of the interval $I_{mn}$.
Because of this, the
potential no longer explicitly appears in the integrand, opening up
the possibility to treat a large class of potentials.  In the limit of
small intervals this becomes exact.  Thus, the replacement
\beq
e^{-i \sum_{n=1}^N V(x_{n})\Delta t}\langle x_N \vert \psi_{i0} (-t)\rangle \to
e^{-i \sum_{n=1}^N V(y_{mn})\Delta t}\langle y_{mN} \vert \psi_{i0} (-t)\rangle .
\label{njm.4}
\eeq
in the integrand of equation (\ref{njm.3}) is expected to be 
a good approximation on
the cylinder set $I_{m_0}\times I_{m_1} \times \cdots \times I_{m_N}$.

Formally the Henstock theory of integration, which is the basis of the 
probabilistic interpretation,  restricts the choice of
intervals, evaluation points and time slices needed for convergence.
However, for smooth short-ranged potentials and wave packets, the
Henstock integrals reduce to ordinary Riemann integrals.  Motivated by
this, it is assumed that convergence can be achieved using
uniformly spaced time slices and intervals of fixed size.  Numerical
convergence provides an indication of the validity of this assumption.
  

The replacement (\ref{njm.4}) leads to the following approximate expression
\beq
\langle x_0 \vert e^{-iHs} \vert \psi_{i0}(-t) \rangle \approx
\lim_{N\to \infty} \lim_{I_{mn}\to 0} 
({ \mu \over 2 \pi i \Delta t})^{N/2}
\sum_{m1 \cdots mN} \prod_{n=1}^N 
(
\int_{I_{mn}}
dx_n
e^{
i {\mu \over 2\Delta t} (x_{n-1}-x_n)^2}
)
e^{-i V(y_{mn}) \Delta t }
\psi_{i0} (y_{mN},-t).  
\label{njm.5}
\eeq

The integrals, 
\beq
P(x_0,I_{m1} \cdots I_{mN}) := 
({\mu \over 2 \pi i \Delta t})^{N/2}
\prod_{n=1}^N  \int_{I_{mn}} dx_n
e^{
i {\mu \over 2\Delta t} (x_{n-1}-x_{n})^2},  
\label{njm.6}
\eeq
are interpreted as complex probabilities to arrive at $x_0$ by passing
through the sequence of intervals $I_{mN} \cdots I_{m1}$. 
The probability interpretation follows because the sum of these quantities
over all intervals is 1, independent of $x_0$:
\beq
\sum_{{m1} \cdots {mN}} 
P(x_0,I_{m1}, \cdots ,I_{mN}) = 1 .
\label{njm.7}
\eeq
This is because, using a simple change of variables, 
the sum can be transformed to the product of $N$ Gaussian-Fresnel 
integrals that are normalized to unity.   

Specifically, $P(x_0,I_{m1}, \cdots ,I_{mN})$ is interpreted as the
``complex probability'' for a path to pass through $I_{mN}$ at time
$t_N$, $I_{mN-1}$ at time $t_{N-1}$, $\cdots$, $I_{m1}$ at time
$t_{m1}$, and end up at $x_0$ at time $t$.  The set of
paths that pass through $I_{mN}$ at time $t_N$, $ \cdots , I_{m1}$ at
time $t_1$ define a cylinder set of paths.  The right most
(initial time) interval only gets contributions from the sample points
$y_{mN}$ that are in the support of the initial wave packet.  Equation
(\ref{njm.7}) is consistent with the requirement that every
path goes through one and only one cylinder set with complex
probability 1.

In \cite{Nelson} Nelson defines a path integral by
analytically continuing the mass in the kinetic energy term.  His
probability is related to the analytic continuation in the mass 
of Muldowney's complex probability.

In this notation the ``path integral'' becomes
\beq
\langle x_0 \vert e^{-iHt} \vert \psi_{i0}(-t) \rangle = 
\lim_{N\to \infty} \lim_{Vol(I_{mn})\to 0}
\sum_{m1 \cdots mN}  P(x_0, I_{m1}, \cdots ,I_{mN})
e^{
-i \sum_{n=1}^{N} V(y_{mn}) \Delta t }
\langle y_{mN} \vert \psi_{i0} (-t)\rangle . 
\label{njm.8}
\eeq
For the half-infinite intervals, $I_{0n}$ and $I_{Mn}$ the upper and
lower boundaries increase (resp. decrease) in the limit.  
Equation
(\ref{njm.8}) is like a Riemann integral with a complex volume element,
except it is interpreted as the expectation of the
random variable
$ e^{ -i \sum_{n=1}^{N} V(y_{mn}) \Delta t }
\langle y_{mN} \vert \psi_{i0} (-t)\rangle$ with
respect to the complex probability distribution
$P(x_0,I_{m1}, \cdots ,I_{mN})$.  Nathanson and J{\o}rgensen show that
the complex probability $P(x_0,I_{m1} \cdots I_{mN})$ is concentrated
on continuous paths and (\ref{njm.8}) converges to a global solution of the
Schr\"odinger equation in the limit of finer partitions and more time
slices.  

This reformulation of Feynman's original path integral provides a
justification to represent time evolution in
quantum mechanics as an average over paths
with complex probabilities.
Transition matrix elements
require an additional multiplication by the potential followed by the
Fourier transform of the resulting quantity.  For the case of equally
spaced sample points this becomes
\[
\int \langle p_f \vert x \rangle V(x) dx 
\sum_{m0,m1 \cdots mN} P(x,I_{m1}, \cdots , I_{mN})
e^{
-i \sum_{n=1}^{N} V(y_{mn}) \Delta t }
\langle y_{mN} \vert \psi_{i0} (-t)\rangle 
\approx
\]
\beq
{1 \over \sqrt{2 \pi}} \sum_{m0, m1, \cdots , mN} e^{-ip_f y_{m0}} \delta y
V(y_{m0}) P(y_{m0},I_{m_1} \cdots I_{m_N}) e^{ -i \sum_{n=1}^{N} V(y_{m_n}) \Delta t }
\langle y_{m_N} \vert \psi_{i0} (-t)\rangle 
\label{njm.9}
\eeq
where $\delta y$ is the width of the $I_{m_N}$ interval, and the 
sum is over the final sample points and the finite intervals, $I_{1_N} 
\cdots I_{M-1,N}$.


  
\section{Factorization}

The input to the Muldowney-Nathanson-J{\o}rgensen formulation of the
path integral is the complex probabilities that a path will be in a
particular cylinder set of paths.  Even if the probabilities
$P(x_0,I_{m_11}, \cdots ,I_{m_NN})$ could be computed analytically,
there are $(M+1)^{N}$ cylinder sets in the limit that $M$ and $N$
become infinite.  Summing over all of these configurations is not
computationally feasible.

On the other hand, for the case of a single time 
step, the same approximations that were made
for multiple time steps lead to the following
\beq
\langle x_{N-1} \vert e^{-iH \Delta t} \vert
\psi_{i0} (-t) \rangle  \approx
\sum_m P(x_{N-1},I_{mN}) e^{
  -i V(y_{mN}) \Delta t }
\langle y_{mN} \vert \psi_{i0} (-t)\rangle . 
\label{f.1}
\eeq
This approximates the transformed wave function after one time step.
The factorization follows if this wave function is used as the initial
state in the transformation to the next time step
\[
\langle x_{N-2} \vert e^{-iH 2\Delta t} \vert \psi_{i0} (-t) \rangle \approx
\sum_{m_{N-1}} P(x_{N-2},I_{m{(N-1)}}) e^{ -i V(y_{m{(N-1)}}) \Delta t
} \langle y_{m(N-1)} \vert e^{-iH \Delta t} \vert \psi_{i0} (-t) \rangle
\approx 
\]
\beq 
\sum_{m(N-1),{m{N}}} P(x_{N-2},I_{m({N-1}}) e^{-i
V(y_{m(N-1)}) \Delta t } P(y_{{(N-1)}} ,I_{mN}) e^{ -i
V(y_{mN}) \Delta t } \langle y_{mN} \vert \psi_{i0} (-t)\rangle . 
\label{f.2}
\eeq
Repeating this for all $N$ time steps gives the following approximation 
\beq
\langle x_0 \vert e^{-i H t} \vert \psi_{i0} (-t) \rangle \approx
\sum_{} P(x_0,I_{m1}) e^{ -i V(y_{m1}) \Delta t} 
\prod_{n=2}^N P(y_{m({n-1})},I_{nm}) e^{ -i V(y_{mn}) \Delta t}
\langle y_{mN} \vert \psi_{i0} (-t)\rangle 
\label{f.3}
\eeq
where the sum is over the cylinder sets.   The factorization
leads to the following approximation of the complex probability
on the cylinder set $\{I_{m1},\cdots , I_{mN}\}$:
\beq
P(x_0,I_{m1},\cdots ,I_{mN}) \approx 
P(x_0,I_{m1}) 
\prod_{n=2}^N P(y_{m({n-1})},I_{nm})
\label{f.4}
\eeq
With this approximation (\ref{s.9}) becomes 
\[
\langle x_0 \vert V   \vert e^{-i H t} \vert \psi_{i0} (-t) \rangle \approx
\]
\beq
\sum_{m1 \cdots mN}
V(x_0) 
P(x_0,I_{m1}) e^{ -i V(y_{m1}) \Delta t} 
\prod_{n=2}^N P(y_{m({n-1})},I_{nm}) e^{ -i V(y_{mn}) \Delta t}
\langle y_{mN} \vert \psi_{i0} (-t)\rangle . 
\label{f.5}
\eeq
This representation has a significant advantage over (\ref{njm.8}) 
because the matrix elements
\beq
K_{m,k} =
P(y_{m},I_{k}) e^{ -i V(y_{k}) \Delta t}
\label{f.6}
\eeq
where
\beq
P(y_{m},I_{k}) = 
({\mu \over 2 \pi i \Delta t})^{1/2}
\int_{x_k}^{x_{k+1}} dx
e^{
i {\mu \over 2\Delta t} (y_{m}-x)^2} 
=
\sqrt{{1  \over i \pi }}
\int_{\sqrt{{\mu \over 2\Delta t}}(x_k-y_m)}^{\sqrt{{\mu \over 2\Delta t}}(
x_{k+1}-y_m)}
e^{ 
i \beta^2} d\beta
\label{f.7}
\eeq
can  be computed analytically,  and  powers  of  this matrix  can  be
computed efficiently.  The integrals in (\ref{f.7}) for finite intervals 
are Fresnel integrals of the form
\beq
I[a,b] = \int_a^b e^{ix^2} dx =
\sqrt{\pi \over 2} (C_{c}(b)- C_{c}(a))
+i \sqrt{\pi \over 2} (S_{c}(b)- S_{c}(a)) .
\label{f.8}
\eeq
where
\beq
C_c(x) = \sqrt{{2 \over \pi}} \int_0^x \cos(t^2) dt
\qquad
S_c(x) = \sqrt{{2 \over \pi}} \int_0^x \sin(t^2) dt .
\label{f.9}
\eeq
Note that these definitions differ from the definitions of
Fresnel integrals given in \cite{Abramowitz} Abramowitz and Stegun.
They are related by
\beq
C_c (\sqrt{{\pi \over 2}}x) = C_{AS} (x)
\qquad
S_c (\sqrt{{\pi \over 2}}x) = S_{AS} (x).
\label{f.10}
\eeq

For the semi-infinite interval with  
$a=-\infty$
\beq
I[-\infty,b] = \int_{-\infty}^b e^{ix^2} dx =
{1 \over 2} \int_{-\infty}^{\infty} e^{ix^2} dx - \int_b^0 e^{ix^2} dx =
\sqrt{{\pi \over 2}} \left ({1+i \over 2} + C_c(b)+i S_c(b) \right ) .
\label{f.11}
\eeq
and for $b=\infty$
\beq
I[a,\infty] = \int_a^{\infty} e^{ix^2} dx =
{1 \over 2} \int_{-\infty}^{\infty} e^{ix^2} dx - \int_0^a e^{ix^2} dx =
\sqrt{{\pi \over 2}} \left ({1+i \over 2} - C_c(a)-i S_c(a) \right ) .
\label{f.11}
\eeq
Using these formulas leads to the following expressions for 
the one-step matrix $K_{mk}$ when $I_k$ is a finite interval:
\[
K_{mk} =
P(y_{m},I_{k})  e^{ -i V(y_{k}) \Delta t} =
\]
\[ 
{1 \over 2}
\left (
(C_{c}(\sqrt{{\mu \over 2\Delta t}}(x_{k+1}-y_m))- 
C_{c}(\sqrt{{\mu \over 2\Delta t}}(x_{k}-y_m)) +
S_{c}(\sqrt{{\mu \over 2\Delta t}}(x_{k+1}-y_m))- 
S_{c}(\sqrt{{\mu \over 2\Delta t}}(x_{k}-y_m)))
\right . 
\]
\beq
\left .
+i (S_{c}(\sqrt{{\mu \over 2\Delta t}}(x_{k+1}-y_m))-
S_{c}(\sqrt{{\mu \over 2\Delta t}}(x_{k}-y_m))- 
C_{c}(\sqrt{{\mu \over 2\Delta t}}(x_{k+1}-y_m))+ 
C_{c}(\sqrt{{\mu \over 2\Delta t}}(x_{k}-y_m)))
\right ) e^{ -i V(y_{k}) \Delta t} . 
\label{f.12}
\eeq
For $x_{k}=x_0= -\infty $:
\[
K_{mk} =
P(y_{m},I_{k0})  e^{ -i V(y_{k}) \Delta t}
=
\]
\beq
{1  \over 2}
\left ( 1 +
(C_{c}(\sqrt{{\mu \over 2\Delta t}}(x_1-y_m))+
S_{c}(\sqrt{{\mu \over 2\Delta t}}(x_1-y_m)))
+i (
S_{c}(\sqrt{{\mu \over 2\Delta t}}(x_1-y_m)) -
C_{c}(\sqrt{{\mu \over 2\Delta t}}(x_1-y_m)))
\right) 
e^{ -i V(y_{0}) \Delta t} 
\label{f.13}
\eeq
and for $x_{k+1}=x_{M+1} \infty $:
\[
K_{mM} =
P(y_{m},I_{M})  e^{ -i V(y_{M}) \Delta t} = 
\]
\beq
{1 \over 2} \left (
1 -( 
S_{c}(\sqrt{{m \over 2\Delta t}}(x_{M}-y_m))+ 
C_{c}(\sqrt{{m \over 2\Delta t}}(x_{M}-y_m))) 
-i(
S_{c}(\sqrt{{m \over 2\Delta t}}(x_{M}-y_m))- 
C_{c}(\sqrt{{m \over 2\Delta t}}(x_{M}-y_m)))
\right )
e^{ -i V(y_{M}) \Delta t} .
\label{f.14}
\eeq
Combining these approximations the expression for 
$\langle \psi_{f0} (0) \vert V \vert e^{-i H \Delta t} \vert \psi_{i0} (-t) \rangle$
is approximately given by
\beq
\langle x \vert  V \vert e^{-i H t} \vert \psi_{i0} (-t) \rangle \approx
\sum_{mk}
V(x)P(x,I_m)e^{-i V(y_m) t}K^{N-1}_{mk}
\langle y_{k} \vert \psi_{i0} (-t)\rangle 
\label{f.15}
\eeq
where $K_{mn}$ is the $(M+1)\times (M+1)$ matrix in equations
(\ref{f.12}-\ref{f.14}).  This matrix only requires $M+1$ values of the
potential.  The other elements needed for this computation are
the minimal uncertainty wave function at time $-t$ at the
same $M+1$ points and the potential.  

The above formulas are for one-dimensional scattering.  The
three-dimensional case involves products of these expressions.
Transition matrix elements can be extracted from (\ref{f.15}) using
(\ref{s.10}).  

\section{Computational considerations}

To test this method, approximate transition matrix elements are
computed for the example of a particle of mass $\mu$ scattering from
an attractive Gaussian potential in one dimension.  These computations
provide a laboratory that can be used to develop strategies for
formulating realistic computations on a quantum computer.

From a mathematical perspective the complex probability interpretation
assumes that all of the integrals are Henstock-Kurzweil integrals.
This means that for given prescribed error, there are restrictions on
how to choose the intervals and evaluation points.  On the 
half-infinite intervals the potential is approximately zero and the
Henstock-Kurzweil integral is a Fresnel integral that can be computed
exactly, while for the finite intervals the Henstock-Kurzweil
integrals are Riemann integrals, so convergence can be realized using
sufficiently fine, uniformly spaced, space and time grids.

In order keep the analysis as simple as possible (1) $N$ time slices
are chosen to be equally spaced and (2) the number, $M-1$, and width
$\delta x$ of the finite intervals on each time slice are chosen to be
identical.

There are a number of constraints that have to be satisfied in order
to get a converged approximation.  The Trotter product formula gives
the exact result in the limit that $(p^2/2\mu)\Delta t$ and
$V\Delta t$ vanish.  In a computation these terms need to be small.
In the first term $p$ is an unbounded operator, but the limit is a
strong limit, so most of the momentum will be centered near the mean
momentum of the initial free particle state.  The potential needs to be
approximately constant on each spatial interval.  The Trotter condition
also requires that  $m \delta x^2 / 2 \Delta t $ is small.
A second constraint is that the uncertainty in
the momentum of the initial state should be less than the momentum.
This avoids having slow moving or backward moving parts of the wave
packet that will feel the potential for long times.  Because of the
uncertainty principle, making $\Delta p$ small makes $\Delta x$ large
- which increases the time that the wave packet feels the potential.
Both $P(x,I_m)$ and $e^{i V(x) \Delta t}$ oscillate, so the widths on
the intervals on each time slice need to be small enough so these
quantities are approximately constant on each interval.  These limits
can be realized by choosing sufficiently small time steps and
sufficiently narrow intervals.  The cost is higher powers of larger
matrices.

Both the range of the potential and spatial width of the initial wave
packet determine the active volume that needs to be broken into 
small intervals.  The velocity of the wave packet determines 
the elapsed time that the initial wave packet interacts with the 
potential.

The condition that ${\mu (\delta x)^2 \over 2 \Delta t}$ is small
requires $N/M^2$ to be small, where $N$ is the number of time steps
and $M$ is the number of intervals per time step.  This means that
shortening the time step may require including more intervals
at each time step.  

For the test an initial wave packet with the dimensionless
parameters used in the calculations are listed in table I.
\begin{table}
\caption{initial wave packet parameters}
\begin{tabular}{ll}
\hline
\hline
mass &$\mu = 1.0$\\
initial momentum &$p_i=   5.0$\\ 
momentum uncertainty &$\Delta p_i =0.25$\\ 
position uncertainty &$\Delta x= 2.0$\\
initial velocity &$v_i=p_i/\mu = 5.0$\\
\hline
\end{tabular}
\end{table}
The particle scatters off of a Gaussian potential of range $r_0$ and strength
$-v_0$,
\beq
V(x) = -v_0 e^{-(x/r_0)^2}.
\label{cc.1}
\eeq
The values of the potential parameters used in the test
calculations are listed in table II.
\begin{table}
\caption{potential parameters}
\begin{tabular}{ll}
\hline
\hline 
strength &$ v0=5.0$\\ 
range &$r0=1.0$\\
\hline
\end{tabular}
\end{table}

The Trotter product formula is justified provided that the time steps
satisfy $(p^2/2m) \Delta t \approx 12.5 \Delta t$ and
$V \Delta t \approx 5 \Delta t$ are small.  The minimum total time is
$t= \mu (r_0+\Delta x)/p = 0.6$.  The calculations require a slightly
longer time for convergence, but convergence can be obtained with a
surprisingly large $\Delta t$.  The active volume is the sum of the
width of wave packet plus the range of the potential, which is about
$3.0$ units.  This must be decomposed into small intervals where the
potential and wave packet are approximately constant.

Graphical methods can be used to determine how long before the collision
the initial wave packet is out of the range of the interaction.
Figure 1 shows a plot of the potential and the initial wave packet at
$t=-3.0$, before it feels the effects of the potential.  The
solid curve represents the potential.  The dashed curve represents the
real part of the initial wave function at $t=-3.0$ and the dotted curve
represents the imaginary part of the initial wave function at $t=-3.0$.
The non-interacting wave packet is evolved to $t=-3.0$ from $t=0.0$ so it
includes the spreading of the wave packet.
This figure suggests that for this initial wave packet and potential that
$\Omega (-\infty) \vert \psi_{i0} \rangle \approx
\Omega (-3.0) \vert \psi_{i0} \rangle $.
%
%
%
%
%
What is relevant
is the combined width of the initial wave packet and the potential
which gives an estimate of the active volume
where the particle feels the effect of the potential. 
Figure 1 indicates that the active volume is about 12
units.  For a wave packet moving with speed $v=p_i/m=-5.0$ the wave
packet will travel 15 units in a time $t=3.0$.  This should be
sufficiently long to move the potential out of the range of the
potential.
%
Figure 1 shows that even with the effects of wave-packet
spreading, at $t=-3.0$ the wave packet has not reached the range of
the potential.  This suggests that $t=3.0$ is a good first guess at
the time $t$ sufficient for convergence of
$\Omega(-t)\vert \psi_{0i}(0) \rangle$.


\begin{figure}
\centering
\caption{Plot of the potential and the incident initial free wave packet
at time $t=-3.0$.}
\includegraphics[scale=.5]{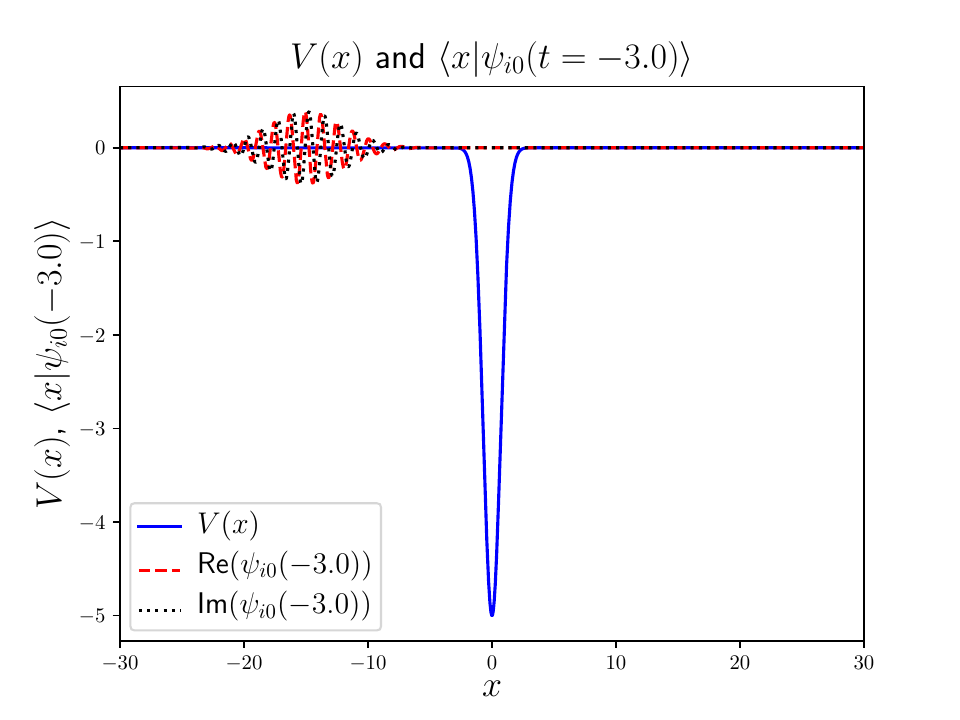}
\label{figure 1}
\end{figure}  

Numerical calculations are possible because of the factorization of
the complex probabilities into products of matrix elements of one-step
probabilities.  Table 3 shows the sum of M=5000 one-step complex
probabilities computed at different points.  The computation shows
that these quantities behave like complex probabilities.  The real
part of the sum of 5000 complex probabilities is always 1.0 and the
imaginary part is always 0, independent of the final $x$ value.  The
table indicates the stability of the sum of these large numbers of
one-step complex probabilities since the cancellation of all of the
imaginary terms is accurate to 15-17 significant figures.
\begin{table}
\caption{Sums of the real and imaginary parts of the complex probabilities.} 
\begin{tabular}{lll}
\hline
\hline
$x$ & $\sum_n Re (P_n(x))$ &$\sum_n Im (P_n(x))$\\
\hline
-25.005001 & p=1.000000$\times 10^0$ & + i 5.273559$\times 10^{-16}$ \\
-20.204041 &p=1.000000$\times 10^0$ &- i 1.387779$\times 10^{-16}$ \\
-15.003001 & p=1.000000$\times 10^0$ & + i 2.775558$\times 10^{-17}$ \\
-10.202040 & p=1.000000$\times 10^0$ &+ i 2.775558$\times 10^{-16}$ \\
-5.001000  & p=1.000000$\times 10^0$ &+ i 7.771561$\times 10^{-16}$ \\
0.200040   &p=1.000000$\times 10^0$ &+ i 3.608225$\times 10^{-16}$ \\
5.001000   &p=1.000000$\times 10^0$ &+ i 1.665335$\times 10^{-16}$ \\
10.202040  &p=1.000000$\times 10^0$ &+ i 5.273559$\times 10^{-16}$ \\
15.003001  &p=1.000000$\times 10^0$ &+ i 8.049117$\times 10^{-16}$ \\
20.204041  &p=1.000000$\times 10^0$&+ i 1.137979$\times 10^{-15}$ \\
25.005001  &p=1.000000$\times 10^0$ &- i 2.164935$\times 10^{-15}$\\
\hline                                          
\end{tabular} 
\end{table} 

The accuracy of the numerical computation of the time evolution of the
initial wave packet depends on having sufficiently small $\delta x$
and $\Delta t$.  The limiting size depends on the initial wave packet.

Since the time evolution of the free wave packet can be computed
analytically, one test of accuracy of the free evolution based on
using products of one-step probabilities is to start with the exact
initial wave packet at $t=-3.0$ and transform it back to the initial
time, $t=0.0$, using multiple applications of the one-step probability
matrices.  This can then be compared to
the exact initial wave packet at $t=0.0$.  In this test $t=-3.0$ is chosen
because the wave packets should be in the asymptotic region
at that time.  This test uses 30 time slices separated by $\Delta_t=0.1$ and
$5000$ spatial steps between $-25.0$ and $25.0$ corresponding to a spatial
resolution $\delta_x=0.01$.  The result of this calculation is
shown in figure 2.  In these plots the dashed lines
represent the calculated $t=0.0$ wave function while the dotted line
represents the exact $t=0.0$ wave function given in (\ref{fs.4}).

These figures compare
\beq
\psi_{i0}(x,0.0) 
\qquad \mbox{to} \qquad
\sum P(x,I_{n1}) P(y_{n1},I_{n2}) \cdots
P(y_{n19 },I_{n20})\langle y_{n20} \vert \psi_{0i}(-3.0) \rangle .
\label{cc.2}
\eeq
In both panels the real and imaginary parts of both wave
functions fall on top of each other.  

\begin{figure}
\caption{Plots of the real and imaginary parts of the free wave packet
at time $t=0$ and the real and imaginary parts of the free path integral
evolved wave packet  
from $t=-3.0$ to $t=0.0$.}
\begin{minipage}[t]{.45\linewidth}
\centering
\includegraphics[scale=.5]{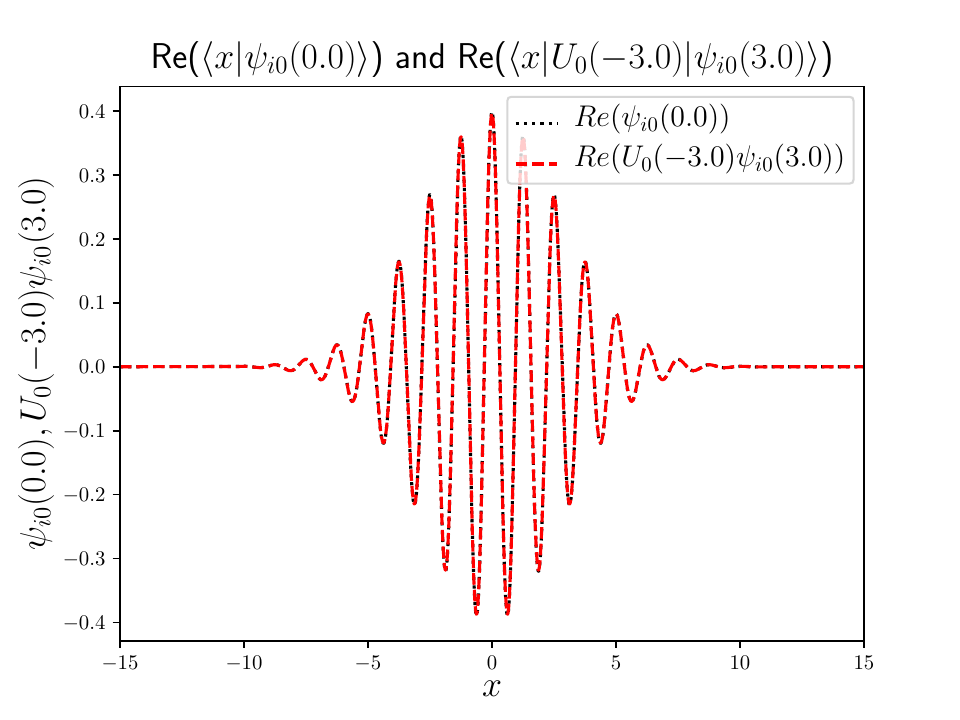}  
\end{minipage}
\begin{minipage}[t]{.45\linewidth}
\centering
\includegraphics[scale=.5]{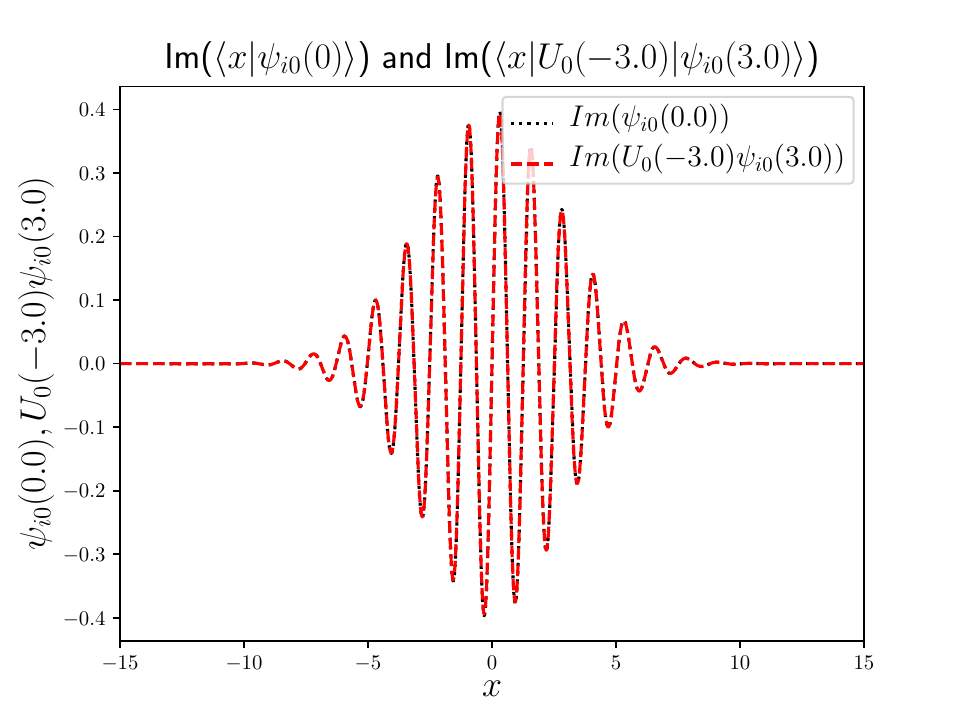}  
\end{minipage}
\label{figure 2}
\end{figure}  
A more detailed examination of the difference appears in figure 3,
which shows the absolute value of the difference of the real and
imaginary parts of the exact and approximate evolved free wave
function divided by the real and imaginary parts of the exact wave
function.  The spikes in the relative error plots are located near the
zeros of the exact wave functions.  The relative error is smallest
between $-4.0$ and $4.0$, which is at the center of the wave packet.
In this region, with the exception of the spike in the imaginary part
at 0, the relative error for the real and imaginary parts of the wave
function in this region is bounded by .007.  Outside of this region
the relative error gets larger.  This is in part because the
path integral has a some fixed numerical uncertainty while   
exact wave function falls off exponentially, so the expression for
the relative error has a denominator that exponentially
approaches 0.    In order to calculate the transition matrix elements
using (\ref{s.10}) or (\ref{s.11}) the exact wave function (with
interaction) only needs to be accurate only in the range of the
potential.

\begin{figure}
\caption{Relative error in the real and imaginary parts of the free path
integral
evolved wave packet
from $t=-3$ to $t=0$.}
\begin{minipage}[t]{.45\linewidth}  
\centering
\includegraphics[scale=.5]{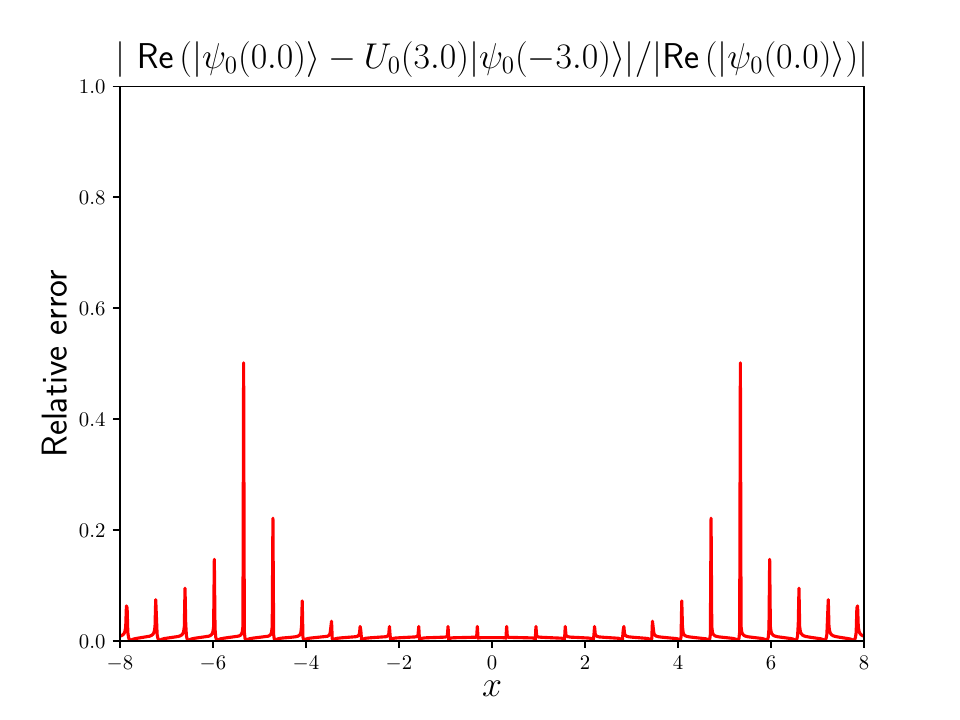}
\end{minipage}
\begin{minipage}[t]{.45\linewidth}  
\centering
\includegraphics[scale=.5]{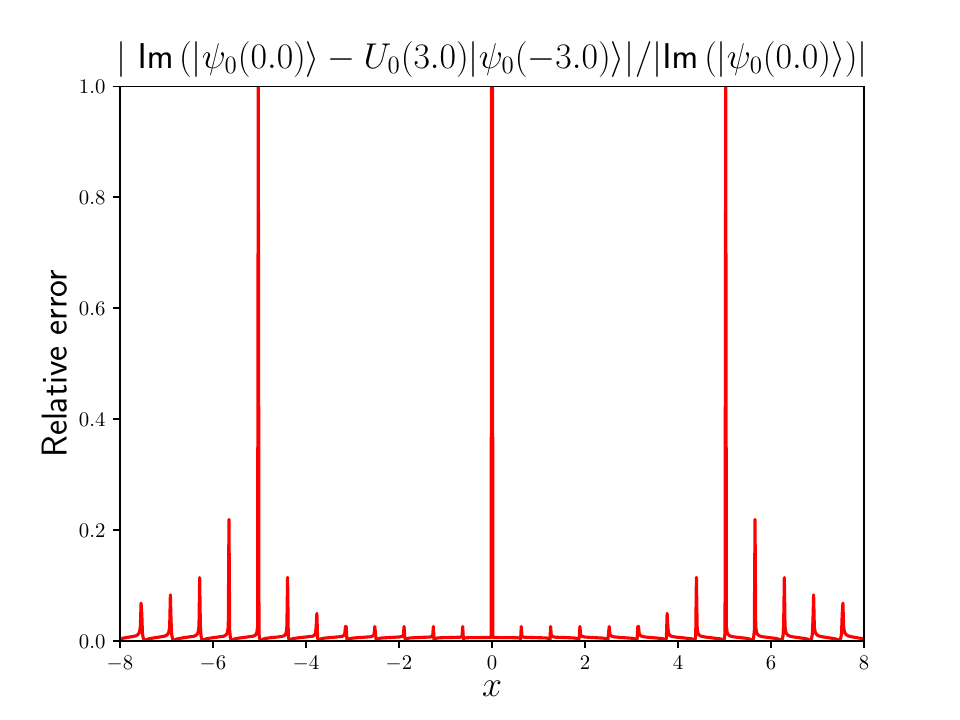}
\end{minipage}
\label{figure 3}
\end{figure}  
These comparisons indicate that both the time steps and resolution are
sufficiently small to evolve the free wave packet for $t=3.0$

In these calculations there was no attempt at efficiency; however
because of the analytic expressions for the one-step
probabilities, the one-step probabilities were computed on the fly in
order to avoid storing large matrices.  The resolution was chosen
to be sufficiently fine to accurately represent the initial wave packet.

The next figure illustrates the effect of the potential on the
evolution of the wave function.  Figure 4 compares the real
and imaginary parts of
the scattering wave function
$\langle x \vert \psi_{i}(0.0)\rangle= \langle x \vert \Omega
(-3.0)\vert \psi_{i0}(0.0)\rangle$ (dotted line) at time $t=0.0$ to the
initial wave packet $\langle x \vert \psi_{i0} (t=0.0)\rangle$ at $t=0.0$.
The parameters used for this computation are $\Delta_t=0.1$,
$\delta_x=0.01$ and $x\in [-25.0,25.0]$, which are the same parameters used
to produce figures 2 and 3.  The functions are calculated at the
midpoint of each finite interval.  Figure 4
shows the change in phase as the interaction is turned on.  The dashed
lines in both panels represent the initial wave packet, while the
dotted lines represent the interacting wave packet.

These calculations approximate the incident scattered
wave function at the collision time $(t=0.0)$.
As discussed in section 2, the exact incident wave function at $t=0.0$
(or any common time) is needed to compute the differential cross section.  These
figures show that both the free and interacting $t=0.0$ wave
functions occupy approximately the same volume, (about 20 units) which
shows that using $\Omega(t)$ to calculate the scattering wave function
eliminates spreading of the wave function.
At time $t=0.0$ the phase of the incident wave is more
dominantly shifted in the forward (right) side of the wave packet, which
has spent the most time interacting with the interaction.
This why these wave
functions do not look like separated transmitted and reflected waves.
\begin{figure}
\caption{Shift in the phase of the real and imaginary parts of the incident wave packet at the collision time $t=0.0$}
\begin{minipage}[t]{.45\linewidth}
\centering
\includegraphics[scale=.5]{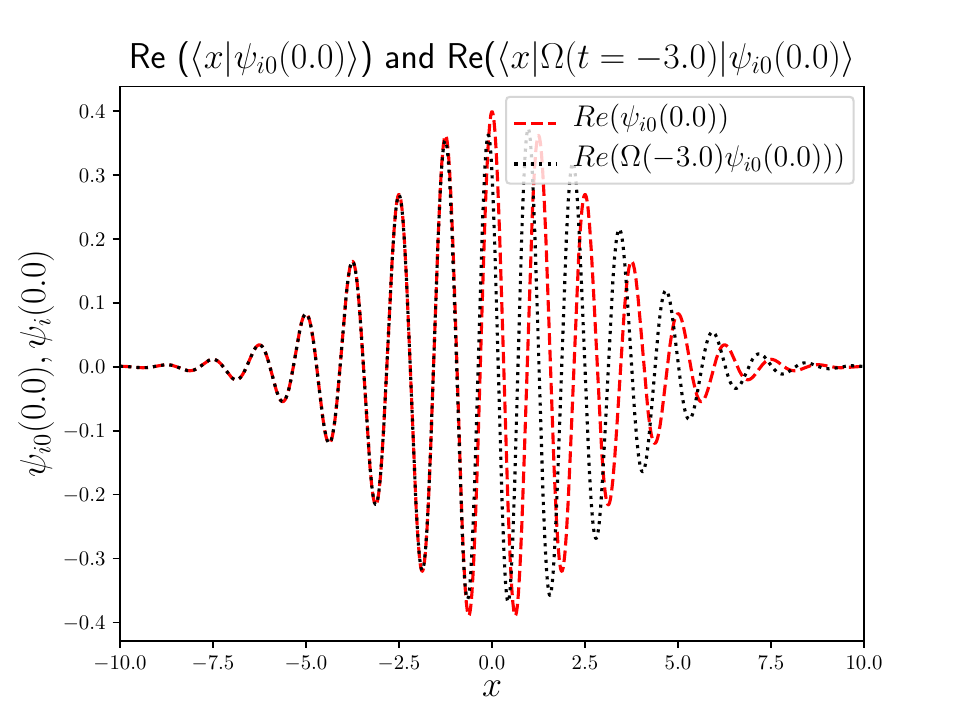}
\end{minipage}
\begin{minipage}[t]{.45\linewidth}
\centering
\includegraphics[scale=.5]{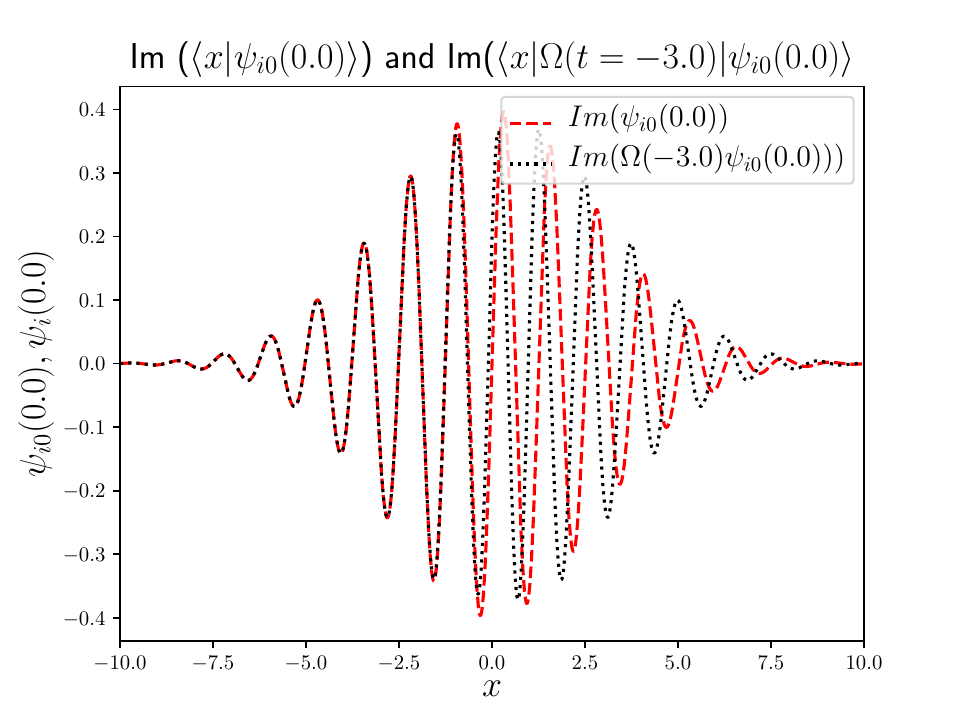}
\end{minipage}
\label{figure 4}
\end{figure}  

One of the complications of performing scattering calculations using
real time path-integral methods is that the final expression for the
transition matrix elements involves several approximations that
need to be tested for convergence.
With the potential turned on it is important to check that (1) the
volume $[-25.0,25.0]$ is sufficiently large, the time step $\Delta t=0.1$
is sufficiently small, the resolution $\delta_x=0.01$ is sufficiently
small, the total time $t=3.0$ is sufficiently large,  and the calculation
is stable with respect to changing the sample point $y_i$ in the
interval $I_i$.  The plots in figures 5-10 investigate the sensitivity 
of the time $t=0.0$ interacting wave functions to variations of the
parameters used in the calculations shown in figure 4.

Note that all of the calculations assume that the evolved wave
functions vanish on the half infinite intervals.  This is justified
both graphically and because the wave packet remains square
integrable.

The Trotter product formula is justified in the small time step limit.
The calculations illustrated in figure 4 used a time step
$\Delta_t=0.1$.  Figure 5 compares the real and imaginary parts of
the scattered wave function $\langle x \vert \psi_{i}(0.0)\rangle= 
\langle x \vert \Omega (-3.0)\vert \psi_{i0}(0.0)\rangle$
using a time step size of $\Delta_t=.1$ (dashed curve) with
corresponding calculations using a time step of half the size,
$\Delta_t=.05$ (dotted curve).  The plots of the real and imaginary
parts of the scattered $t=0$ wave functions for $\Delta_t=.1$ and
$\Delta_t=.05$ fall on top of each other.  This indicates that the
time resolution $\Delta_t$ is sufficiently fine for this calculation
\begin{figure}
\caption{Plots of the real and imaginary parts of the time $t=0$
scattered wave packet
calculated using Trotter time steps of $\Delta t =0.1$ and
$\Delta t =0.05.$}
\begin{minipage}[t]{.45\linewidth}
\centering
\includegraphics[scale=.5]{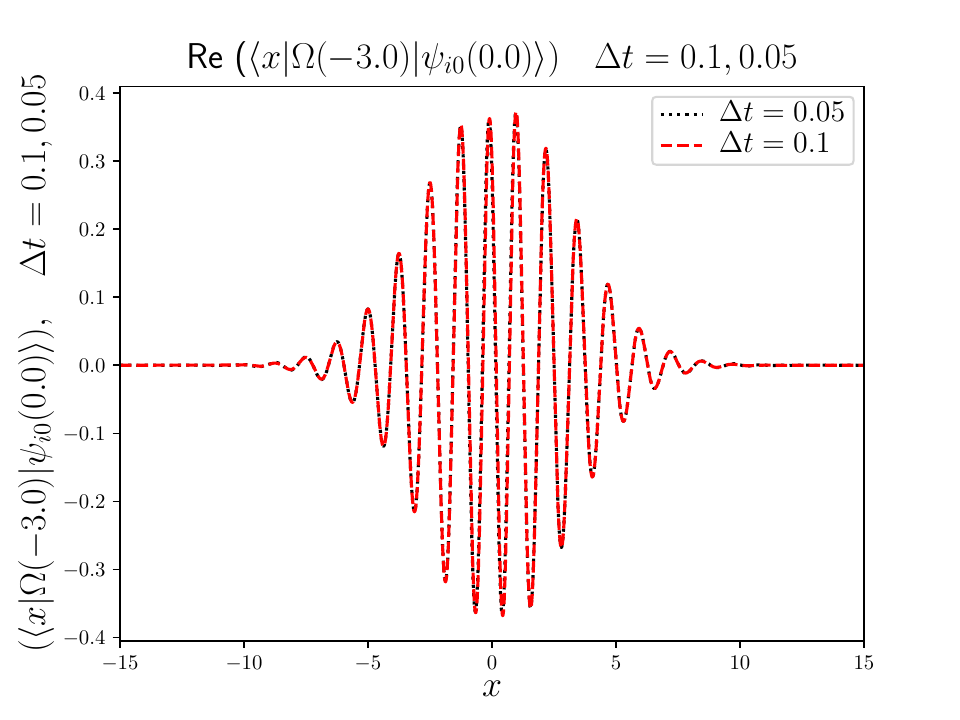}
\end{minipage}
\begin{minipage}[t]{.45\linewidth}
\centering
\includegraphics[scale=.5]{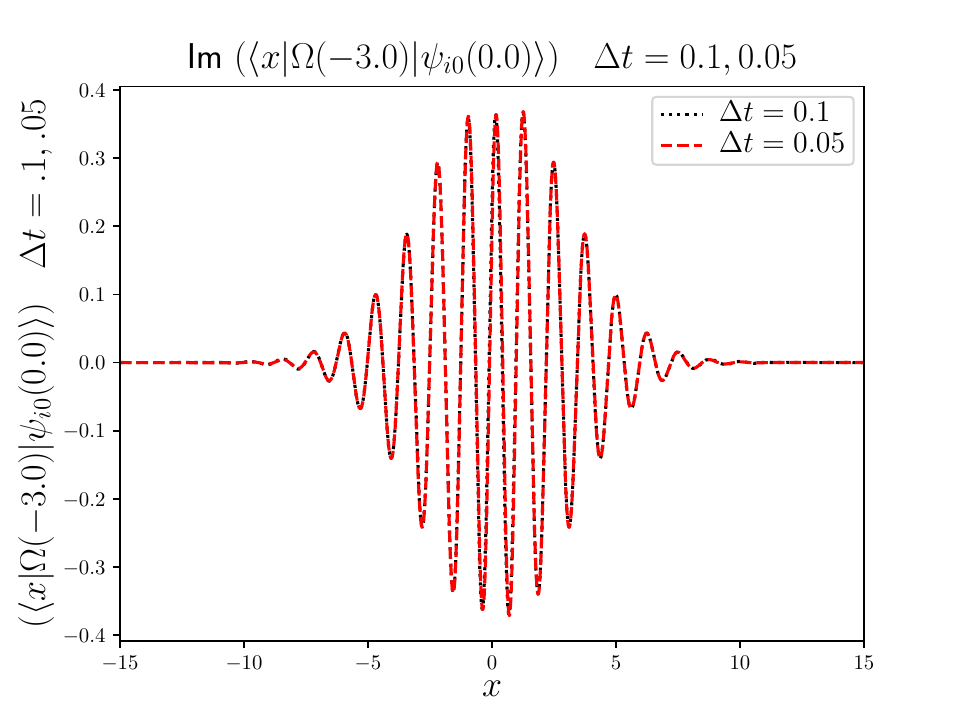}
\end{minipage}
\label{figure 5}
\end{figure}  

While the Trotter approximation requires a sufficiently small time
step, an accurate approximation of the wave operators requires a
sufficiently large time.  The initial choice of the approximating
$\Omega_- := \lim_{t \to -\infty}\Omega(t)$ by $\Omega (-3.0)$ was
determined by examining the range of the potential and width and speed
of the wave packet.  Graphical methods indicated that the incident wave
packed was still in the asymptotic region at $t=-3$.  
Figure 6 shows the
effect of increasing the time from $t=3.0$ to $t=6.0$, keeping the
size of the time step $\Delta_t=0.1$ constant, on the real and
imaginary parts of the calculated scattering wave function.  The plots
of the wave functions for $t=-3.0$ (dashed curve) and $t=-6.0$ (dotted
curve) fall on top of each other.  This suggests that $t=3.0$ is
sufficient for convergence.

\begin{figure}
\caption{Plots of the real and imaginary parts of the scattered wave packet
using approximate wave operators evaluated at $t=-3.0$ and $t=-6.0$.} 
\begin{minipage}[t]{.45\linewidth}
\centering
\includegraphics[scale=.5]{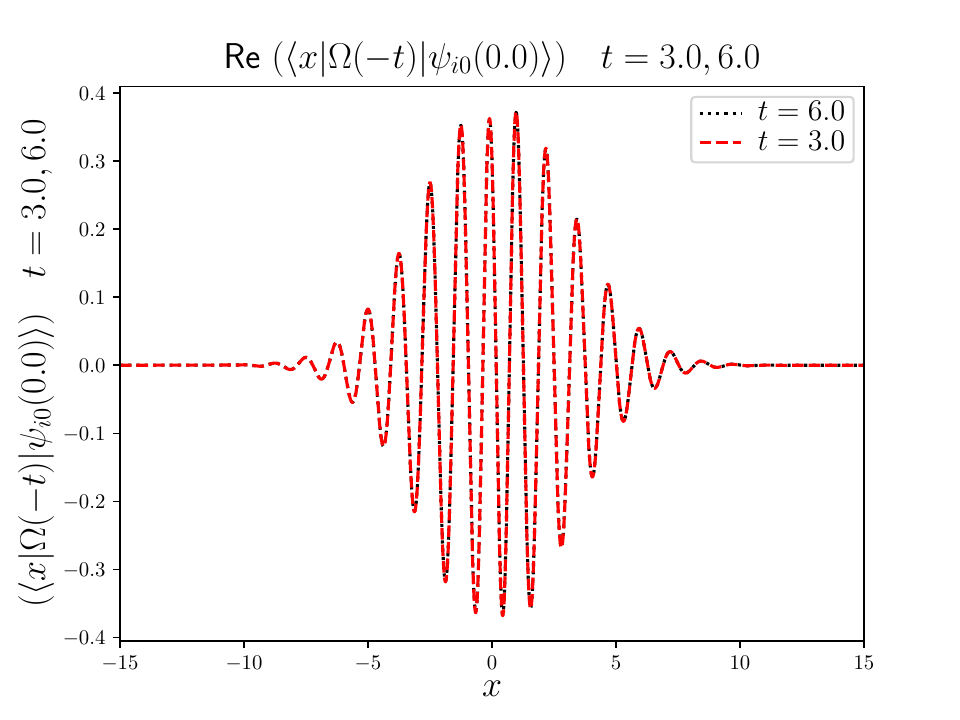}
\end{minipage}
\begin{minipage}[t]{.45\linewidth}
\centering
\includegraphics[scale=.5]{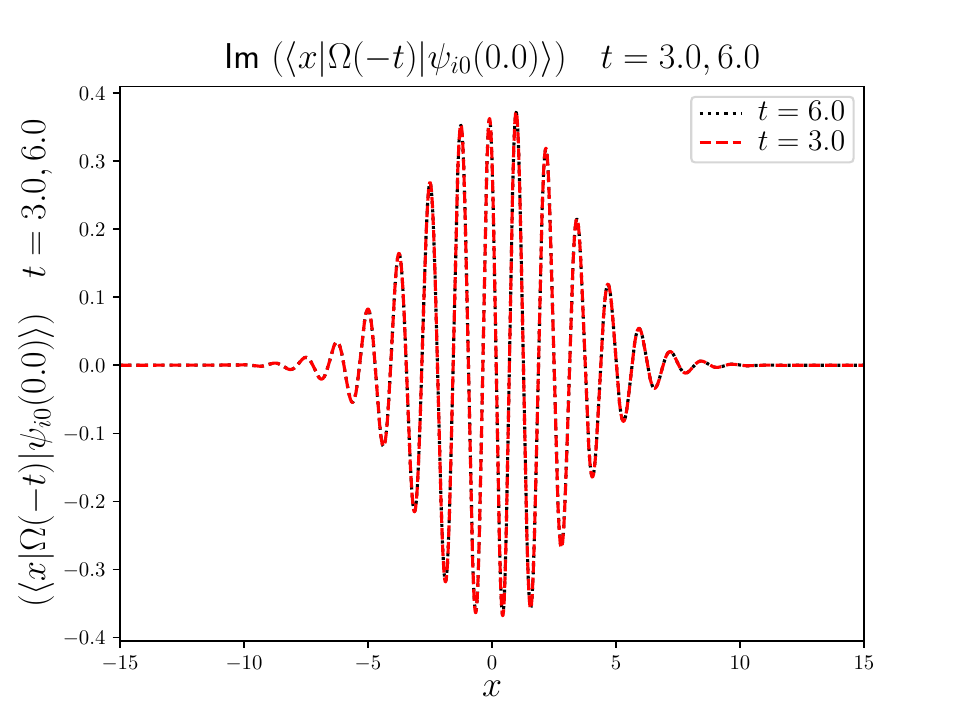}
\end{minipage}
\label{figure 6}
\end{figure}  

The other limit used in this formulation of the path integral is the
decomposition of the volume at each time slice into sufficiently small
intervals. The
spatial resolution of the intervals that define the cylinder sets
should be sufficiently small that the wave function is effectively
constant on them.  The required resolution must be small 
due to the oscillating nature of the wave function,
Figure 7 shows the effect of increasing the
spatial resolution from 5000 intervals ($\delta_x=0.01$, dashed curves)
to 10000 intervals ($\delta_x=0.005$, dotted curves) on the real and
imaginary parts of the scattering wave function.  In these
calculations $\Delta_t=0.1$, $t=3.0$ and the interval is $[-25.0,25.0]$.
Again the curves for the real and imaginary part of the wave function
fall on top of each other.  This indicates that for this problem a
spatial resolution $\delta_x=.01$ (5000 intervals)) is sufficient for
convergence.

\begin{figure}
\caption{Plots of the real and imaginary parts of the scattered
wave packet using cylinder sets with spacial resolutions of $0.01$ and $0.005$.}
\begin{minipage}[t]{.45\linewidth}
\centering
\includegraphics[scale=.5]{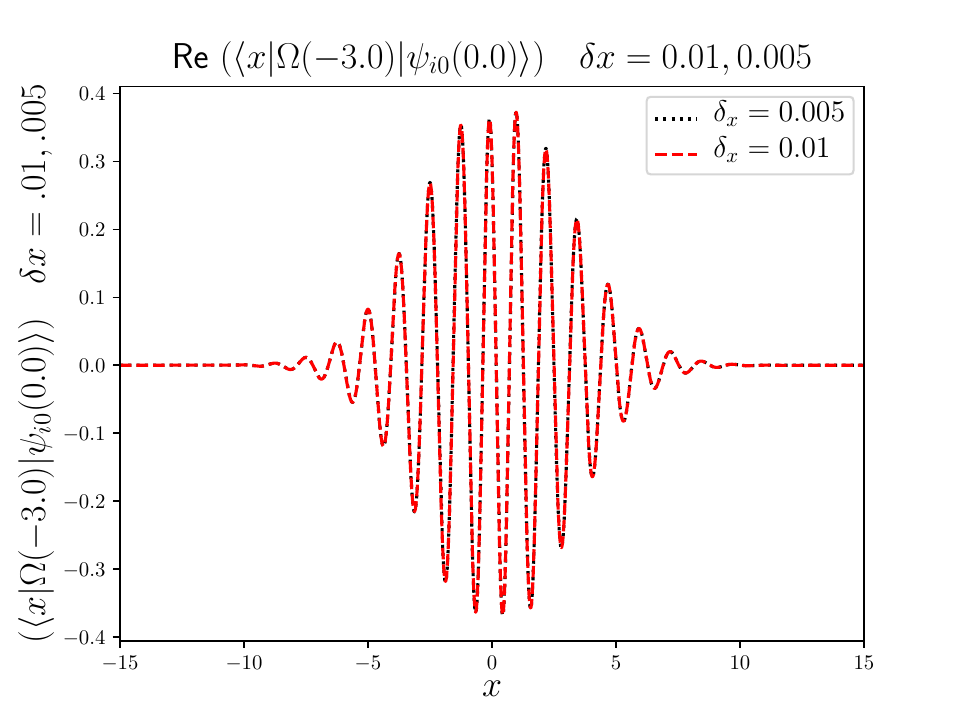}
\end{minipage}
\begin{minipage}[t]{.45\linewidth}
\centering
\includegraphics[scale=.5]{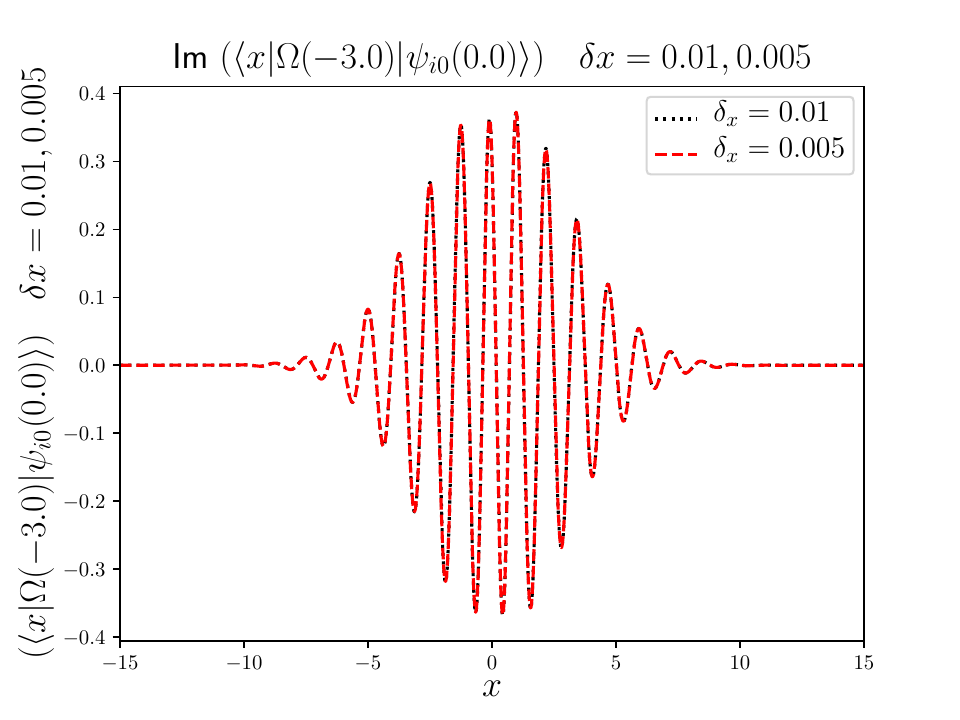}
\end{minipage}
\label{figure 7}
\end{figure}  

While the previous plots suggest that wave functions vanish outside of
the volume $[-25.0,25.0]$, it is still important to check that the results
are stable with respect to increasing the active volume of the
calculation.  Figure 8 compares calculations of the real and
imaginary parts of the wave function where the volume is changed from
$[-25.0,25.0]$ (dotted curves) to $[-50.0,50.0]$ (dashed curves) keeping
$\delta_x=0.01$, $\Delta_t=0.1$, and $t=3.0$ Again, the calculations
indicate the volume $[-25.0,25.0]$ is sufficient for convergence.

\begin{figure}
\caption{Plots of the real and imaginary parts of the scattered wave packet
computed in the regions $[-25.0,25.0]$ and $[-50.0,50.0]$.}
\begin{minipage}[t]{.45\linewidth}
\centering
\includegraphics[scale=.5]{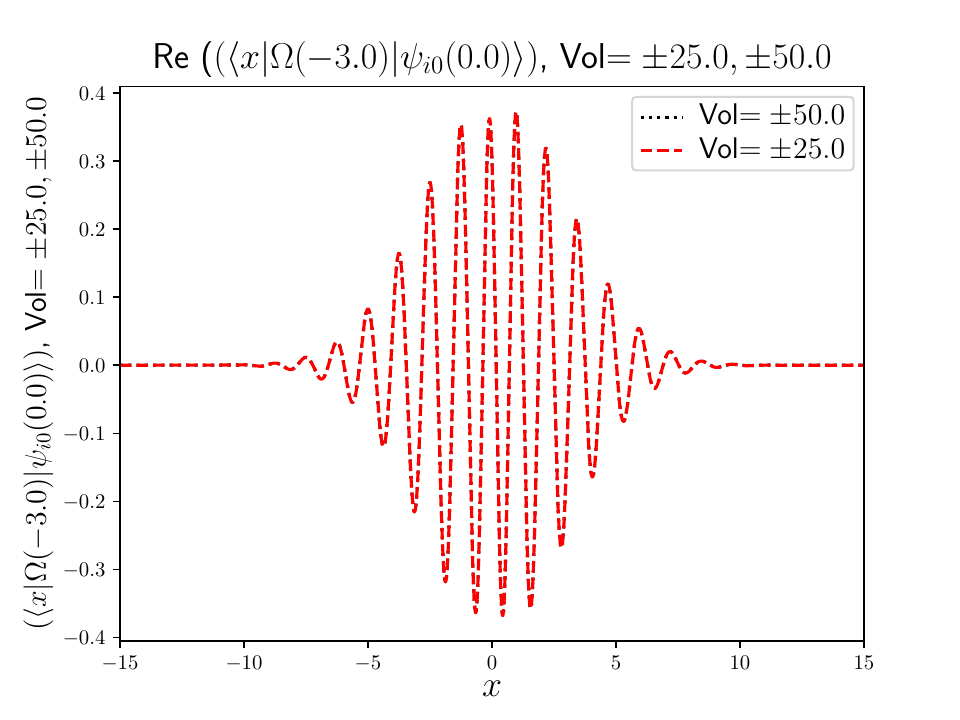}
\end{minipage}
\begin{minipage}[t]{.45\linewidth}
\centering
\includegraphics[scale=.5]{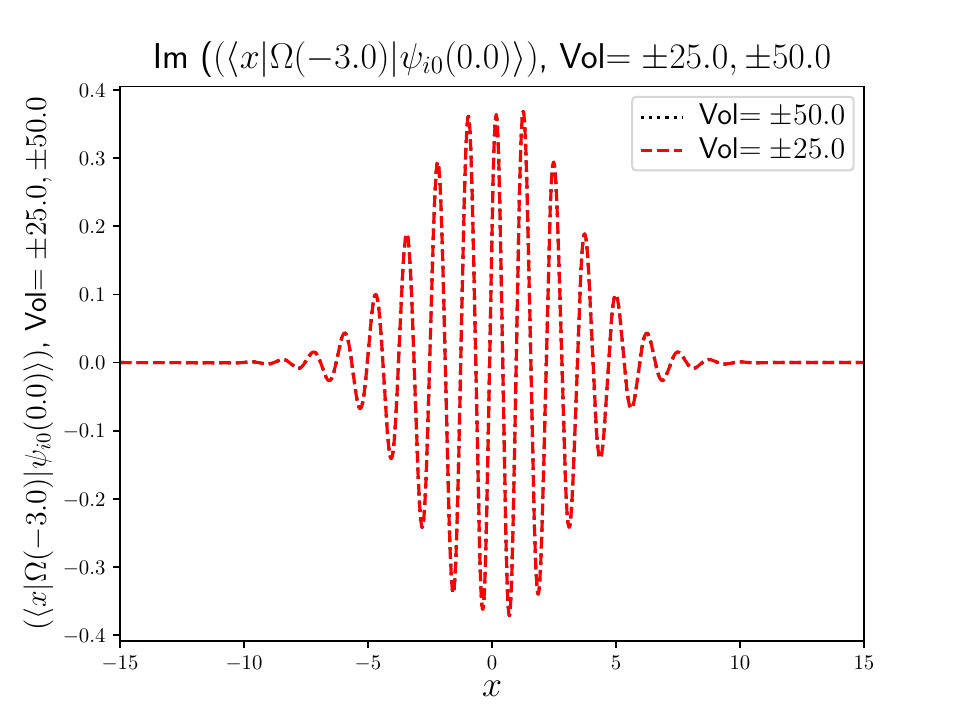}
\end{minipage}
\label{figure 8}
\end{figure}

The size of the spatial intervals, $\delta x$, should be sufficiently
small that the potential and initial wave free wave packet are
approximately constant on the intervals.  Figure 9 compares
calculations where the potential is evaluated at the center or left
endpoint of each interval for $\delta x=0.01$. Figure 9 shows the real
and imaginary parts of the wave function, where the sample points are at
the left
(dash-dot line) and the center (dashed line) of the interval.
This tests whether the potential is
locally constant on each interval.  These graphs show a small shift in
the overall phase of the wave function.  Figure 10 repeats the
calculations shown in Figure 9 by increasing the resolution
by a factor of 2.  These figures show a corresponding reduction in the
difference between the two curves.  The conclusion is that the
calculations in figure 4 seem to be most sensitive to the choice of
the position in each interval used to evaluate the potential.  This
sensitivity may be due to large slope
near the edges of the Gaussian potential shown in figure 1. 
\begin{figure}
\caption{Plots of the real and imaginary parts of the scattered wave packet
with the potential evaluated the at the center or left end of each
interval for $\delta x=0.01$. 
}
\begin{minipage}[t]{.45\linewidth} 
\centering
\includegraphics[scale=.5]{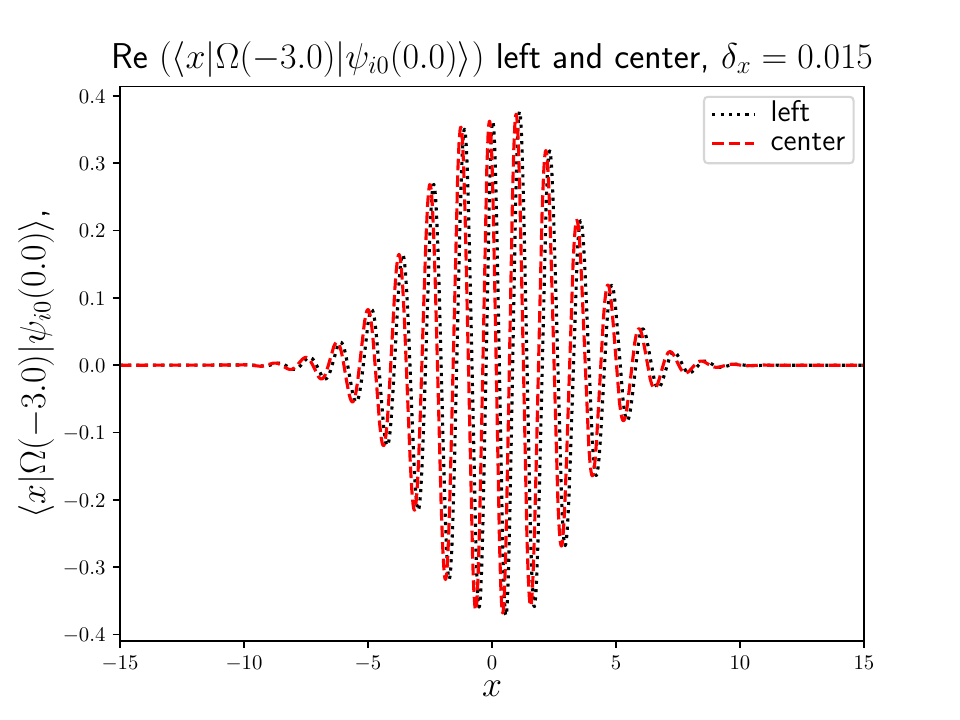}
\end{minipage}
\begin{minipage}[t]{.45\linewidth}
\centering
\includegraphics[scale=.5]{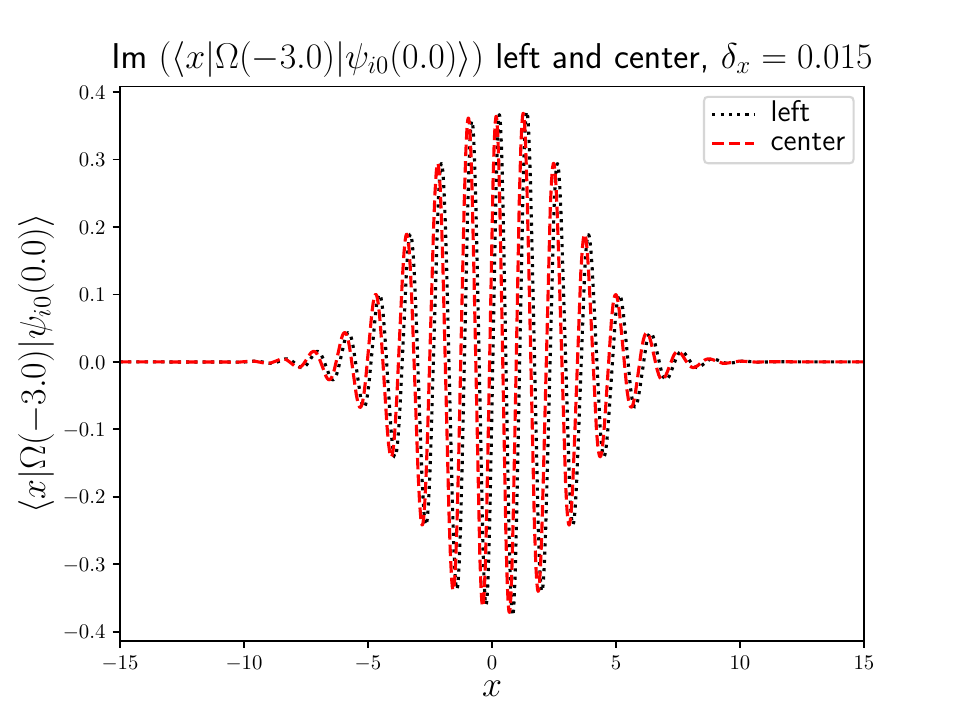}
\end{minipage}
\label{figure 9}
\end{figure}
 
\begin{figure}
\caption{Plots of the real and imaginary parts of the scattered wave packet
with the potential evaluated the at the center or left end of each
interval for $\delta x=0.005$.
}
\begin{minipage}[t]{.45\linewidth} 
\centering
\includegraphics[scale=.5]{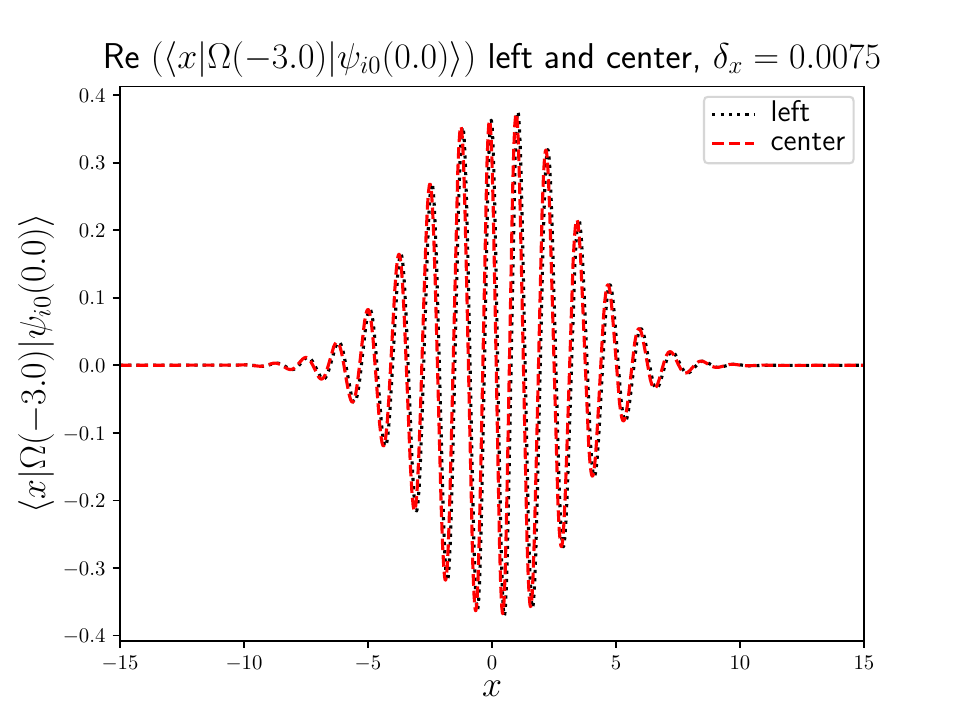}
\end{minipage}
\begin{minipage}[t]{.45\linewidth}
\centering
\includegraphics[scale=.5]{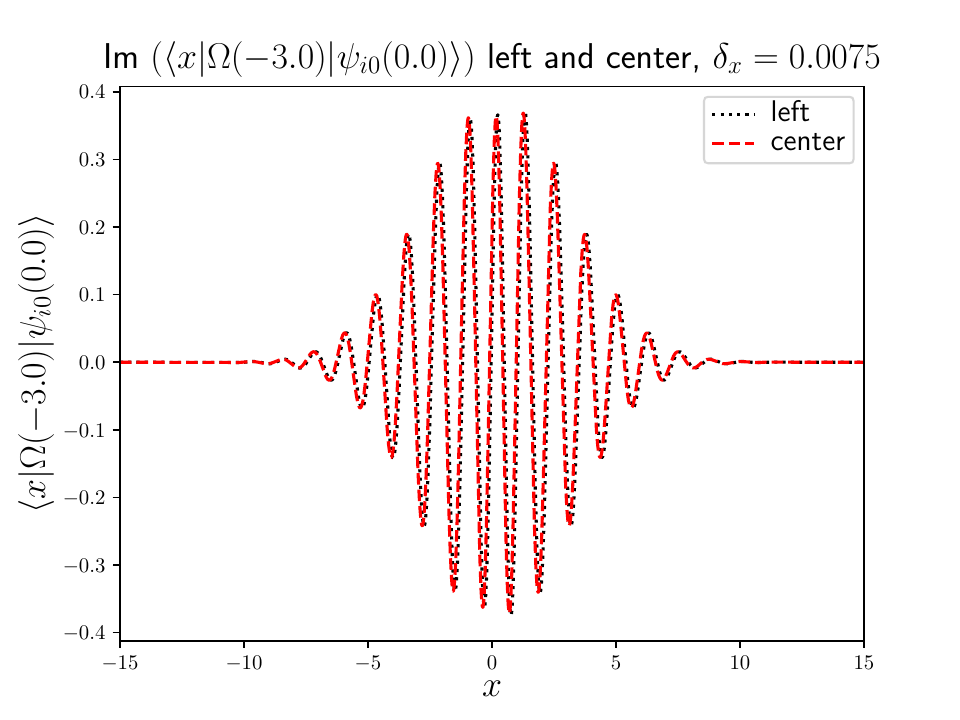}
\end{minipage}
\label{figure 10}
\end{figure}

The goal of this work is to determine if these scattering wave
functions can be used to calculate sharp-momentum transition matrix
elements.  Because the limits in the Trotter product formula are
strong limits, the initial momentum was replaced by a narrow wave
packet.  To test the effect of the smearing, the sharp-momentum Born
approximation is compared to the Born approximation where the initial
sharp momentum state is replaced by a Gaussian delta function of width
.25, which was used in the calculations above.

This comparison is illustrated in figure 11.  The solid curve shows
the Gaussian approximation to a delta function with width $\Delta p=.25$.
The dotted curve shows the potential with initial momentum $p=5.0$,
as a function of the final momentum.  The dashed curve shows the 
same potential with the initial momentum replaced by the Gaussian 
delta function state centered at $p=5.0$ as a function of the final momentum. 
\begin{figure}
\caption{Plots of the delta function normalized Gaussian,
the exact Born approximation to $T$, and the Born approximation
smeared with the initial Gaussian wave packet.}
\centering
\includegraphics[scale=.5]{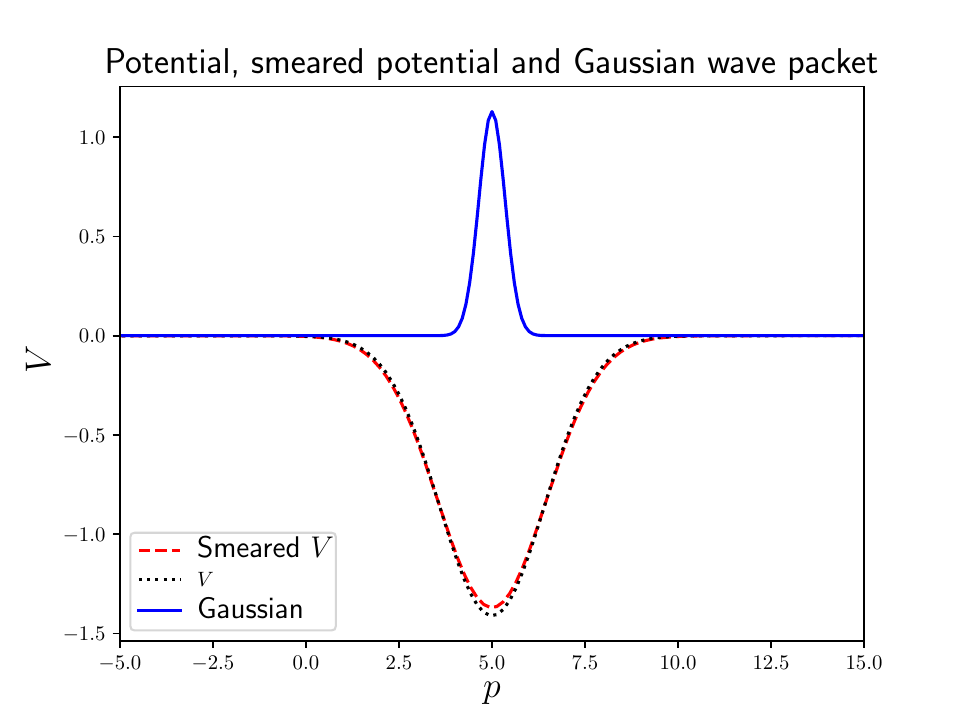} 
\label{figure 11}
\end{figure}  
The figure shows a small decrease in the matrix elements due to 
the smearing near the on-shell value.
 
As another test of the numerical convergence,  the smeared Born 
approximation $\langle p \vert V \vert \Psi_{i0}(0)\rangle$ is compared to 
$\langle p \vert V \Omega_0 (V=0,t=-3.0) \vert \Psi_{i0}(0.0)\rangle$, where
the {\it potential is turned off} to compute $\Omega_0(-3.0)$.  
In this calculation 
$\delta x= 0.01$ and $\Delta t =0.1$.  This comparison is shown in figure
12.  The imaginary part of $\langle p \vert V \Omega_0 (V=0,t=-3.0) \vert
\Psi_{i0}\rangle$, shown in the solid line should vanish, while the
real parts of 
$\langle p \vert V \vert \Psi_{i0}(0.0)\rangle$ (dotted line) and 
$\langle p \vert V \Omega_0 (V=0,t=-3.0) \vert \Psi_{i0}(0.0)\rangle$
(dashed line), should agree.  The figure shows that 
the calculation accurately approximates the smeared Born approximation.

\begin{figure}
\caption{Plots of the exact smeared Born approximation,
the real and imaginary parts of the Born approximation smeared
against the wave packet
computed using complex probabilities with V set to 0.} 
\centering
\includegraphics[scale=.5]{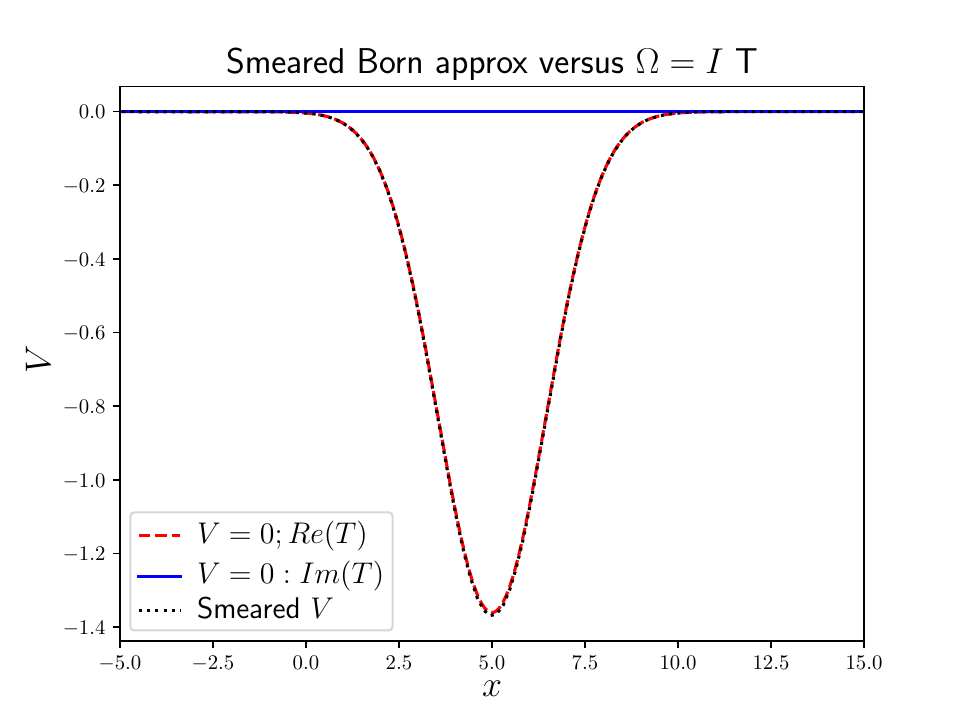}
\label{figure 12}
\end{figure}  
The approximate calculations give
$V(x) \langle x \vert \Omega (-t) \vert \psi_{i0}\rangle$.  The
sharp-momentum transition matrix elements can be computed using a
direct Fourier transform (\ref{s.10}) or by integrating against
Gaussian delta function with the desired final momentum (\ref{s.11}).
These two methods of calculation are compared in figure 13.
In these calculations the $\Delta p$ of the final Gaussian is 0.25,
which is the same value used in the initial Gaussian.  The dashed
curves show the real and imaginary parts of the smeared transition
matrix elements computed using a direct numerical Fourier transform
(dashed curves) compared to the curves which show the corresponding
quantities that replace the final momentum by a Gaussian approximation
to a delta function (dotted curves).  For these calculations the time
step was taken to be $\Delta t =0.025$, and $\delta_x= 0.003$ which is
smaller than the time step used in the calculations shown in figures 4-10.
This is because the $T$-matrix calculations, which involve
Fourier transforms or integration against oscillating wave packets,
are more sensitive to the accuracy of the scattering wave functions.
These figures show that both methods give results within a few percent
of each other.

\begin{figure}
\caption{Plots of the real and imaginary parts of the $p=5.0$ half-shell T matrix computed by numerical Fourier transform and
by smearing with a delta-function normalized Gaussian wave packet.
}
\begin{minipage}[t]{.45\linewidth}
\centering
\includegraphics[scale=.5]{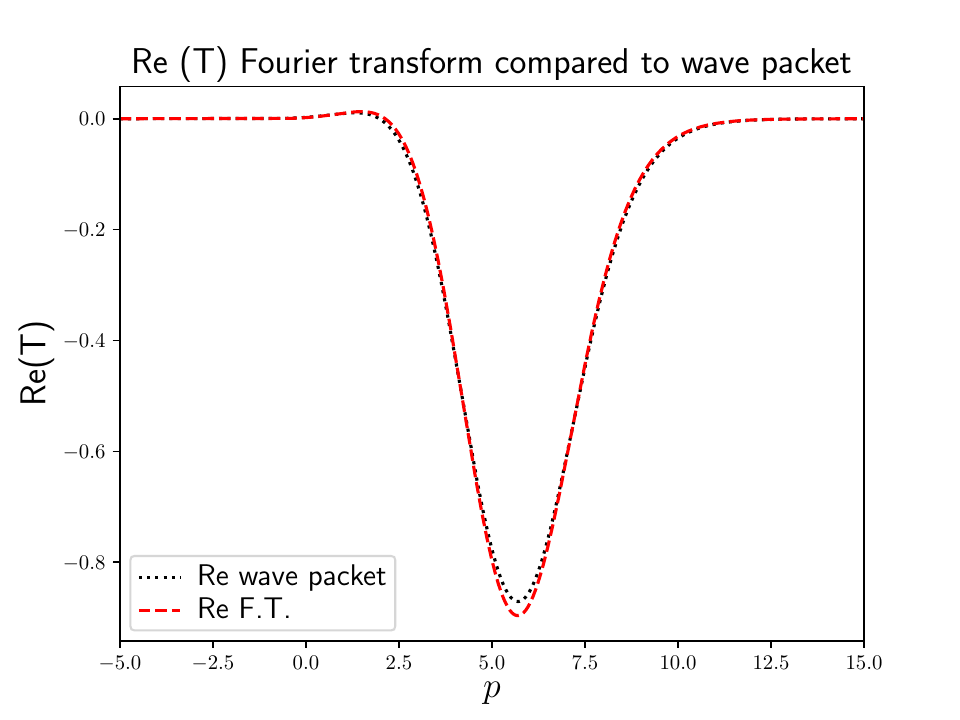}
\end{minipage}
\begin{minipage}[t]{.45\linewidth}
\centering
\includegraphics[scale=.5]{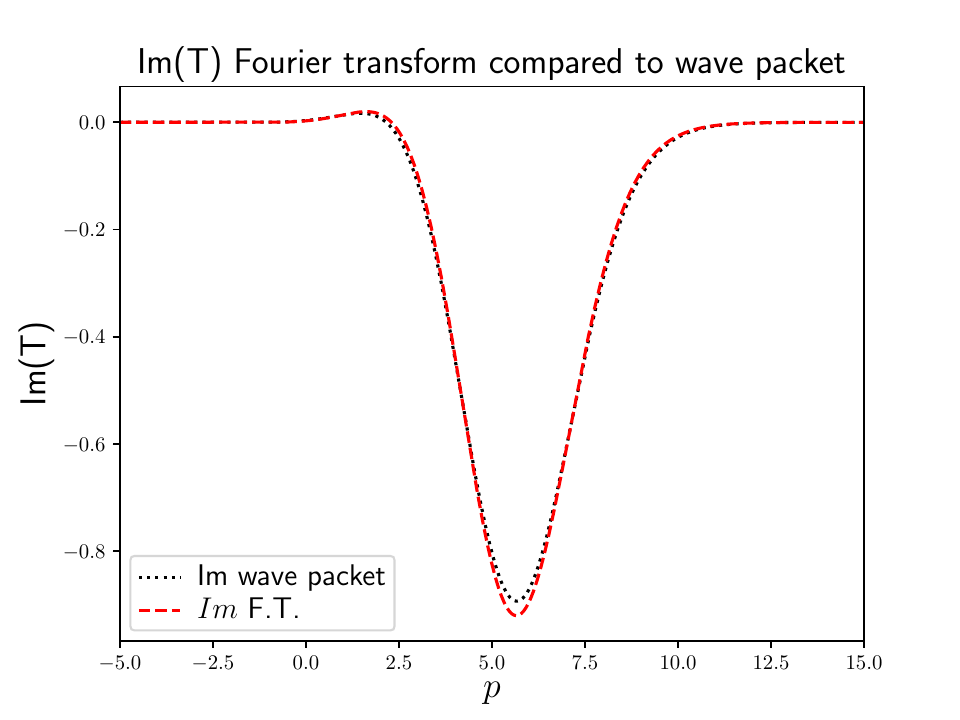}
\end{minipage}
\label{figure 13}
\end{figure}  

Figure 14 compares the real and imaginary parts of the smeared
half-shell transition matrix elements (using the direct Fourier
transform method (dotted curves) to sharp momentum half-shell
transition matrix elements computed by numerically solving the
Lippmann-Schwinger equation (dashed curves).
As with the other calculation, errors
are a few percent at the on-shell point.  The comparison is between a
sharp-momentum matrix element and a matrix element where the initial
state is smeared with a narrow wave packet, so there will be some
residual difference due to the smearing.  The relative error
for the real and imaginary parts are illustrated in figure 15.
While the relative errors can be large, they are a few percent near the
on shell point, $p=5.0$.  All of the errors can
be reduced by increasing the number of time steps and intervals.
 
\begin{figure}
\caption{
Plots of the real and imaginary parts of the
half-shell transition
matrix computed using the Fourier transform of the path integral
calculation and by numerical solution of the Lippmann-Schwinger
equation ($p=5.0$).}
\begin{minipage}[t]{.45\linewidth}
\centering
\includegraphics[scale=.45]{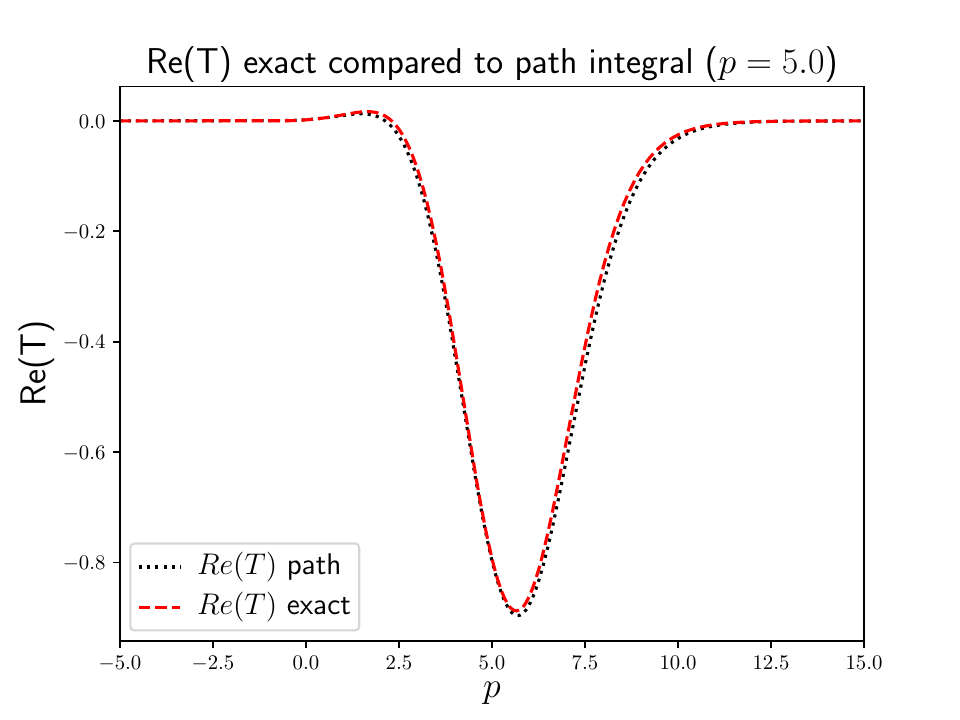}
\end{minipage}
\begin{minipage}[t]{.45\linewidth}
\centering
\includegraphics[scale=.45]{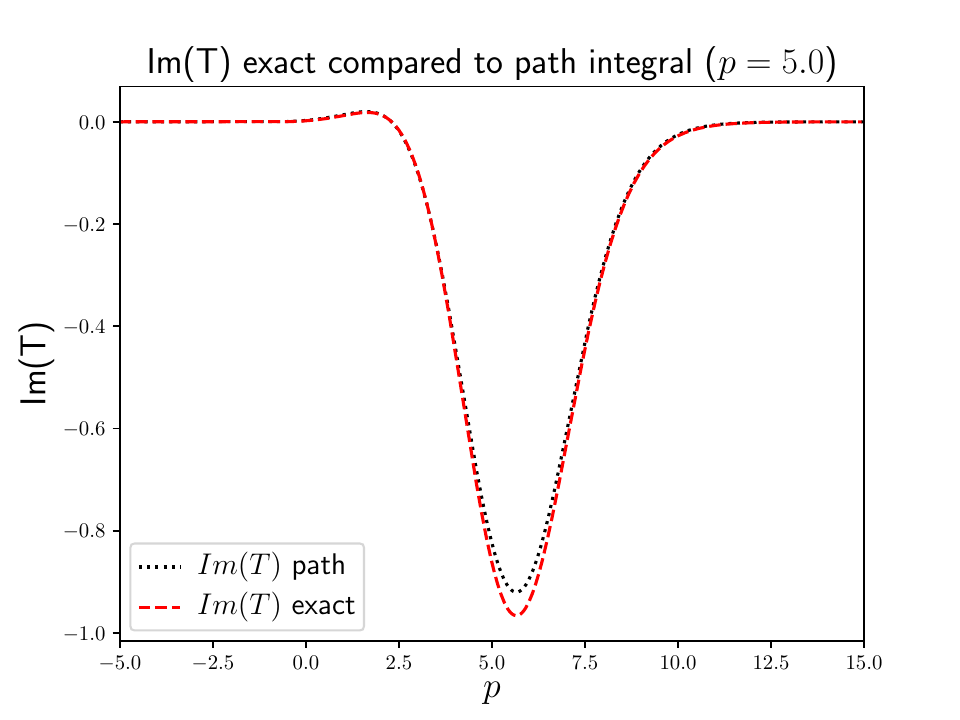}
\end{minipage}
\label{figure 14}
\end{figure}

\begin{figure}
\caption{Plot of the absolute value of the relative error
of the real and imaginary parts of the half-shell transition matrix
path integral calculated using path integrals and
the Lippmann-Schwinger equation where $p=5.0$ is the on-shell point.}
\begin{minipage}[t]{.45\linewidth}  
\centering
\includegraphics[scale=.5]{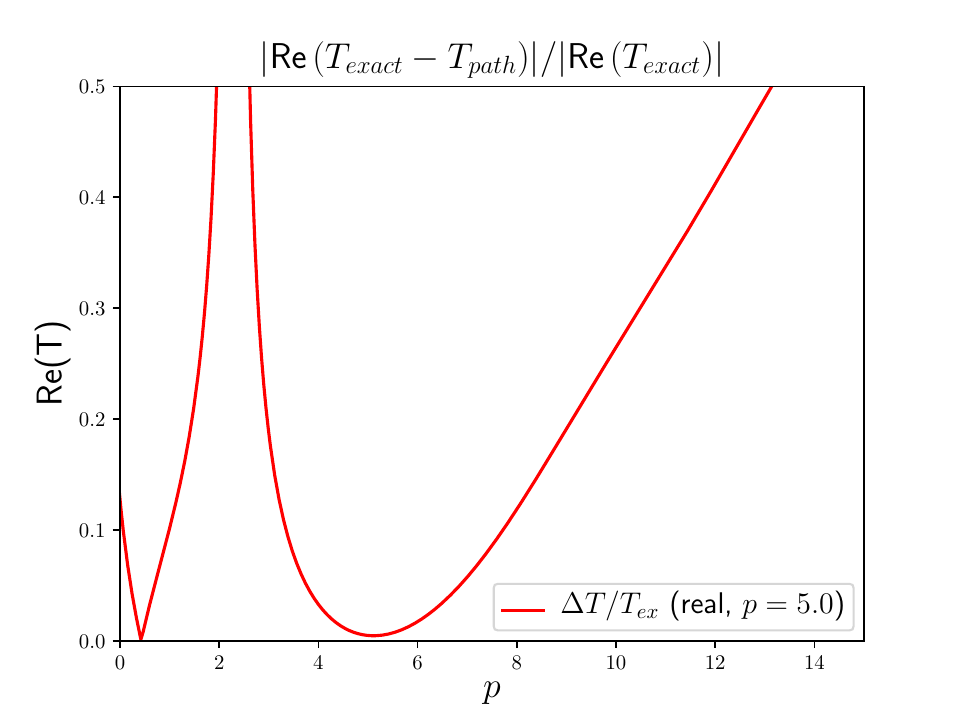}
\end{minipage}
\begin{minipage}[t]{.45\linewidth}
\centering
\includegraphics[scale=.5]{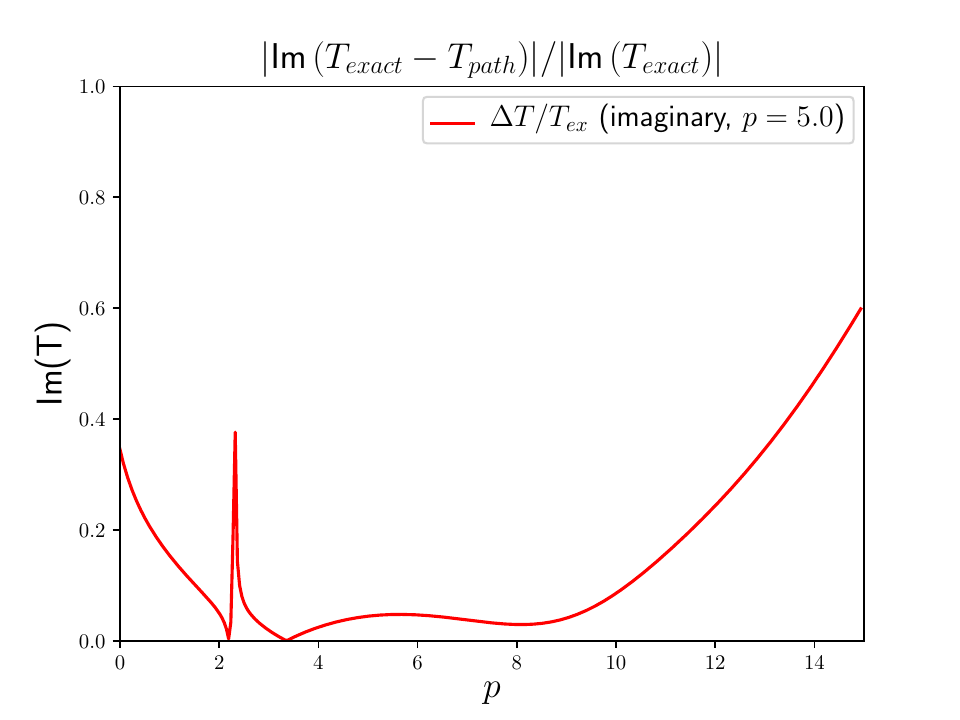}
\end{minipage}
\label{figure 15}
\end{figure}  

In this one-dimensional problem there are two on-shell transition
matrix elements corresponding to the transmitted wave and reflected
wave.  The calculations exhibited in the series of figures above are
only for the transmitted wave.  The on-shell matrix element for the
reflected wave can be evaluated by changing the sign of the final
momentum either in the Fourier transform on in the Gaussian
approximation to the delta function.  At this energy the transition
matrix elements for the reflected wave are about five orders of
magnitude smaller that those of the transmitted wave and are not
reliable.

Phase shifts, transmission and reflection coefficients can be
computed from the transition matrix elements.  The results for
$\Delta t=0.025$, $\delta_x= 0.003$ between $-15.0$ and $15.0$,
$\Delta p =0.25$ are shown below. 
For the Fourier Transform method
the transition matrix elements are
\[
\langle p \vert T \vert p \rangle = -8.04\times 10^{-01} -i 8.17 \times 10^{-01}  \qquad
\langle -p \vert T \vert p \rangle = - 2.41 \times 10^{-06} + i3.16 \times 10^{-06} .
\]
The phase shift for forward scattering is 
\[
\delta (p) =23.0^{\circ} .
\]
A numerical calculation of the Lippmann-Schwinger equation
gives a phase shift of $\delta (p)= 25.4^{\circ}$
which is slightly larger than the computed phase shift.
The phase shift for the reflected wave involves a ratio of two small
numbers, both of which are smaller than computational errors,
so it was not computed. 

The transmission and reflection coefficient are
\[
T=-0.027 + i 1.01  \qquad 
R=  3.03 \times 10^{-6} +i 3.99 \times 10^{-6}
\]
and the unitarity check gives
\[
\vert T\vert ^2+\vert R\vert ^2= 1.02
\]
For the Gaussian approximate delta function 
the transition matrix elements are
\[
\langle p \vert T \vert p \rangle = -7.85e^{-01} -i 7.96e^{-01} \qquad
\langle -p \vert T \vert p \rangle = -2.16e^{-06}  +i 3.35e^{-06} .
\]
and the phase shifts are
\[
\delta(p) = 23.0^{\circ} \qquad \delta(-p) = 0.0^{\circ} 
\]
which agree with the phase shifts calculated using
the Fourier transform method.
The reflection and transmission coefficients are
\[
T=-1.59\times 10^{-04} +i 0.986 \qquad
R =2.71\times 10^{-06} +i 4.2 \times 10^{-06}\\
\]
and the unitarity check gives
\[
\vert T\vert ^2+\vert R\vert^2= 0.972
\]
Note that there are slight differences in the unitary checks for
the two methods of computation.

For the calculations at $p=5.0$, the momentum was sufficiently high that
the amplitude of the reflected wave was negligible relative to the
transmitted wave.

Two additional calculations were performed at lower momenta ($p=2.5$
and $p=1.0$) that have larger amplitude reflected waves.  For these
calculations the width of the initial wave packet in momentum had to
be sufficiently small so the spreading in position would not outrun
the motion of the center of the wave packet.  Decreasing the momentum
width of the initial wave packet increases the spatial volume of the
wave packet.  This, along with the slower moving wave packet,
increases the time that the particle feels the potential.

The results of these calculations are summarized in table III.  The
calculations in the table used the direct Fourier transform method,
although it gives the same phase shifts as the Gaussian wave packet
method.  The calculations at $p=1.0$ used $15000$ spatial intervals
between $-35.0$ and $35.0$ with a spatial resolution of
$\delta_x=0.0047$.  The total time was $t=45.0$ which was broken up
into and $1000$ time steps of size $\Delta t=0.045$.  The width of the
initial wave packet was $\Delta p= 0.1$.

The calculations at $p=2.5$ used $16000$
spatial intervals between $-15.0$ and $15.0$ with a spatial resolution of
$\delta_x=.0019$.  The total time was $t=7.0$ which was broken up into
and $1000$ time steps of size $\Delta t= 0.007$.  The width of the
initial wave packet was $\Delta p= 0.2$.  The phase shifts and unitarity
checks for the three calculations are shown in table 2

\begin{table}
\caption{Phase shifts and unitarity checks}
\begin{tabular}{llllll}
\hline
\hline
$p$ & $\phi^t$ & $\phi^t_e$ & $\phi^r$ & $\phi^r_e$ & $R^2+T^2$ \\
\hline
1   & -49.1$^{\circ}$ & -44.9$^{\circ}$  & 0.4$^{\circ}$ & 0.1$^{\circ}$ & 0.7\\
2.5 & 82.7$^{\circ}$ &82.9$^{\circ}$    & -74.8$^{\circ}$  & -52.1$^{\circ}$ & 0.9\\ 
5.  & 23.0$^{\circ}$ & 25.4$^{\circ}$ & 0.0$^{\circ}$ & 0.0$^{\circ}$ & 1.0 \\
\hline
\end{tabular}
\end{table}

The half-shell transition matrix elements for these two momenta are
shown in figures 16-19.  The real and imaginary parts of the
calculated $p=1.0$ half-shell transition matrix elements are compared
to the same quantities by numerically solving the Lippmann-Schwinger
equation in figure 16.  The path integral calculation reproduces the
qualitative features of the solution of the Lippmann Schwinger
equation.  The corresponding relative errors for these two
calculations are shown in figure 17.  The relative errors are between
0.1 and 0.2 near the on-shell point.

These lower energy calculations required larger volumes and more time steps than
the calculation at $p=5.0$, resulting in a much slower convergence.
\begin{figure}
\caption{Plots of the real and imaginary parts of the
half-shell transition
matrix computed using the Fourier transform of the path integral
calculation and by numerical solution of the Lippmann-Schwinger
equation ($p=1.0$).}
\begin{minipage}[t]{.45\linewidth} 
\centering
\includegraphics[scale=.5]{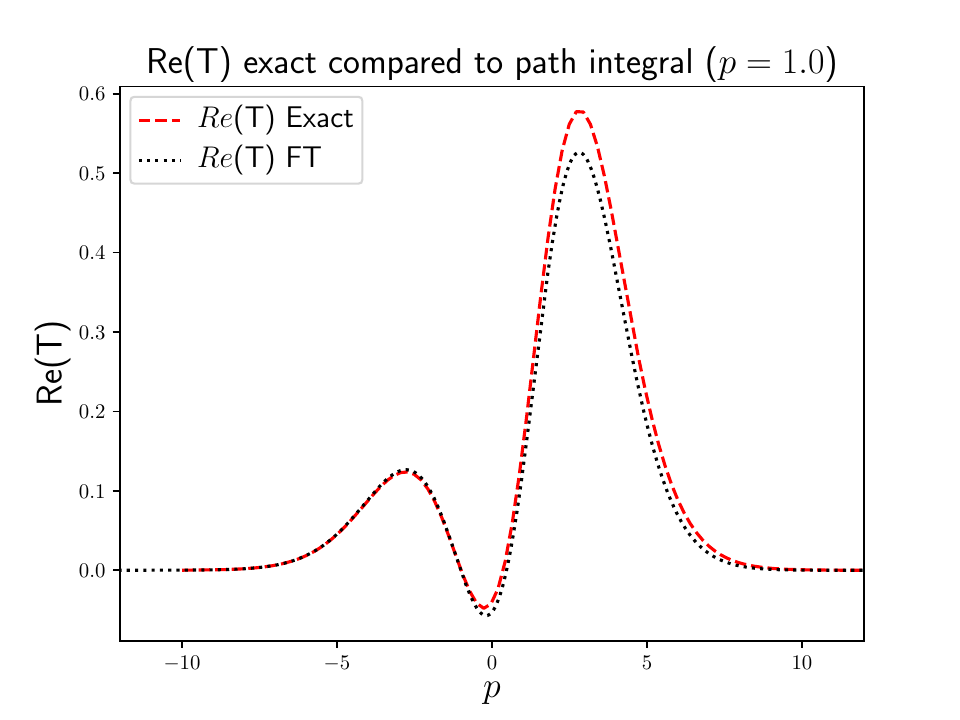}
\end{minipage}
\begin{minipage}[t]{.45\linewidth}
\centering
\includegraphics[scale=.5]{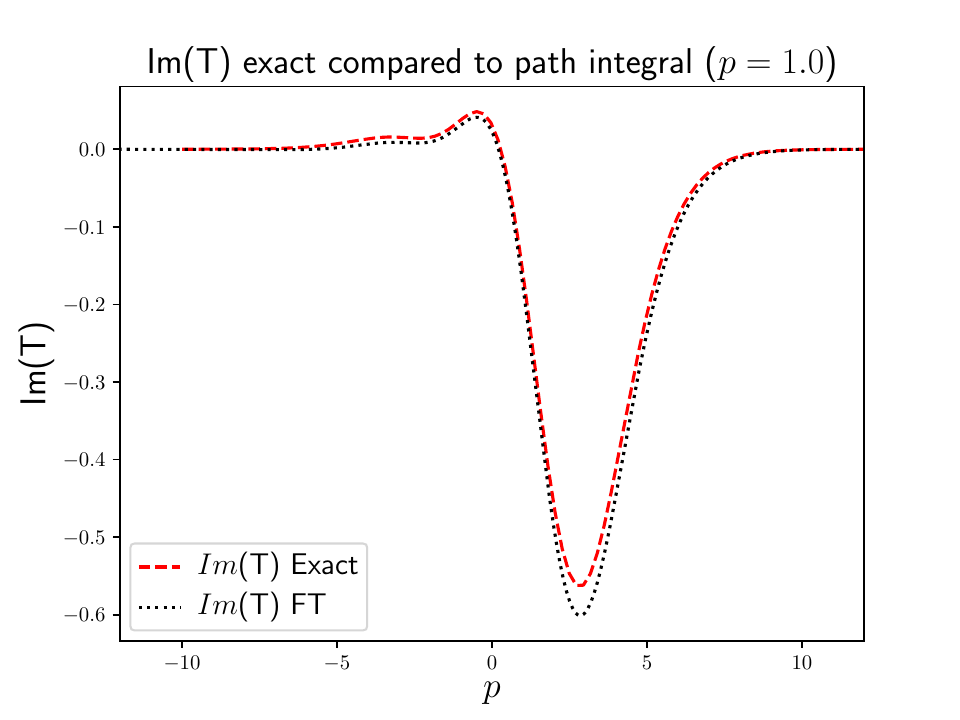}
\end{minipage}
\label{figure 16}
\end{figure}

\begin{figure}
\caption{Plot of the absolute value of the relative error
of the real and imaginary parts of the half-shell transition matrix
path integral calculated using path integrals and
the Lippmann-Schwinger equation where $p=1.0$  is the on-shell point.
} 
\begin{minipage}[t]{.45\linewidth}  
\centering
\includegraphics[scale=.5]{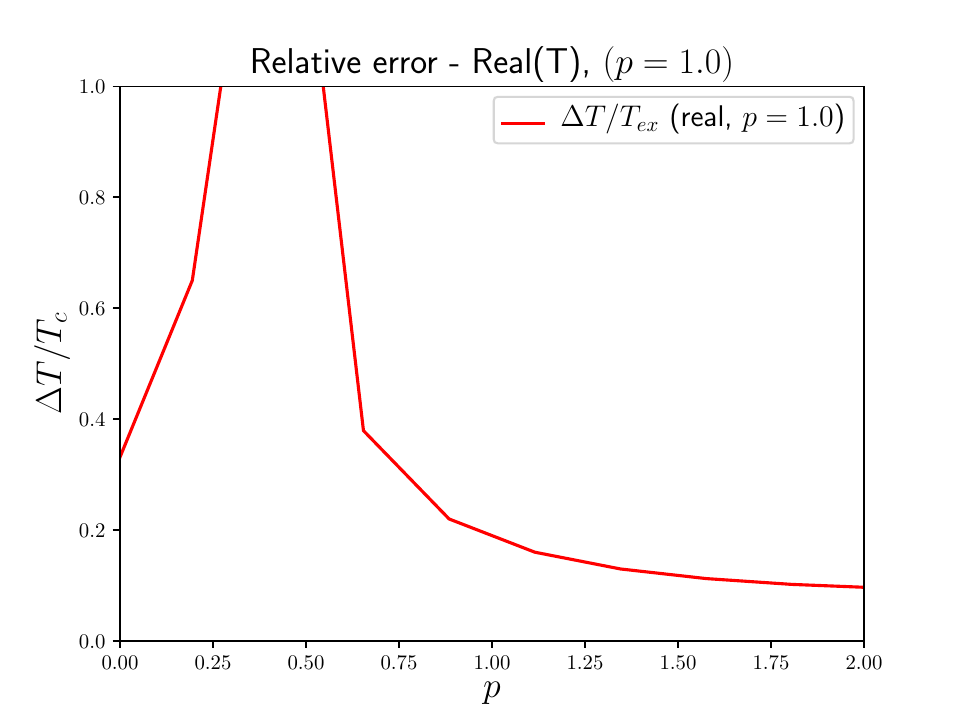}
\end{minipage}
\begin{minipage}[t]{.45\linewidth}  
\centering
\includegraphics[scale=.5]{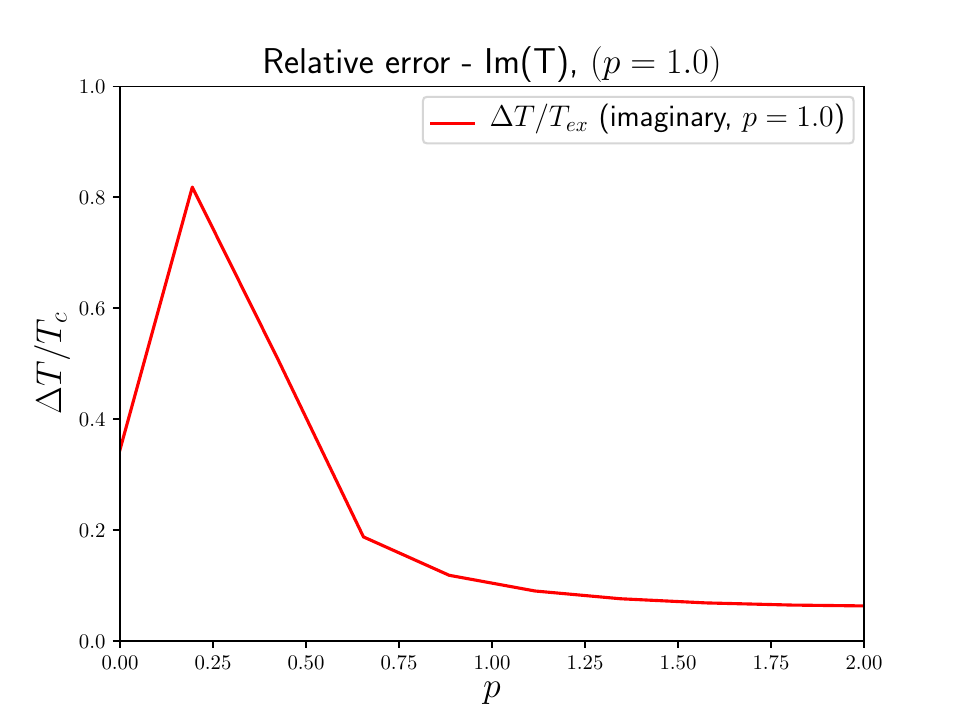}
\end{minipage}
\label{figure 17}
\end{figure}  
The real and imaginary parts of the half-shell transition matrix
elements for $p=2.5$ computed with the path integral and Lippmann
Schwinger equation are compared in figures 18 and 19.  The real part
of the transition matrix element is almost a factor of 10
smaller than the imaginary part.  In addition, from the
graph, near the on shell point $p=-2.5$ the real and imaginary parts
of the transition operator are near 0 so the calculation of the phase
shift, which for the reflected wave is a function of the ratio of
these two small quantities, is dominated by computational errors.

\begin{figure}
\caption{Plots of the real and imaginary parts of the
half-shell transition
matrix computed using the Fourier transform of the path integral
calculation and by numerical solution of the Lippmann-Schwinger
equation ($p=2.5$).}
\begin{minipage}[t]{.45\linewidth}  
\centering
\includegraphics[scale=.5]{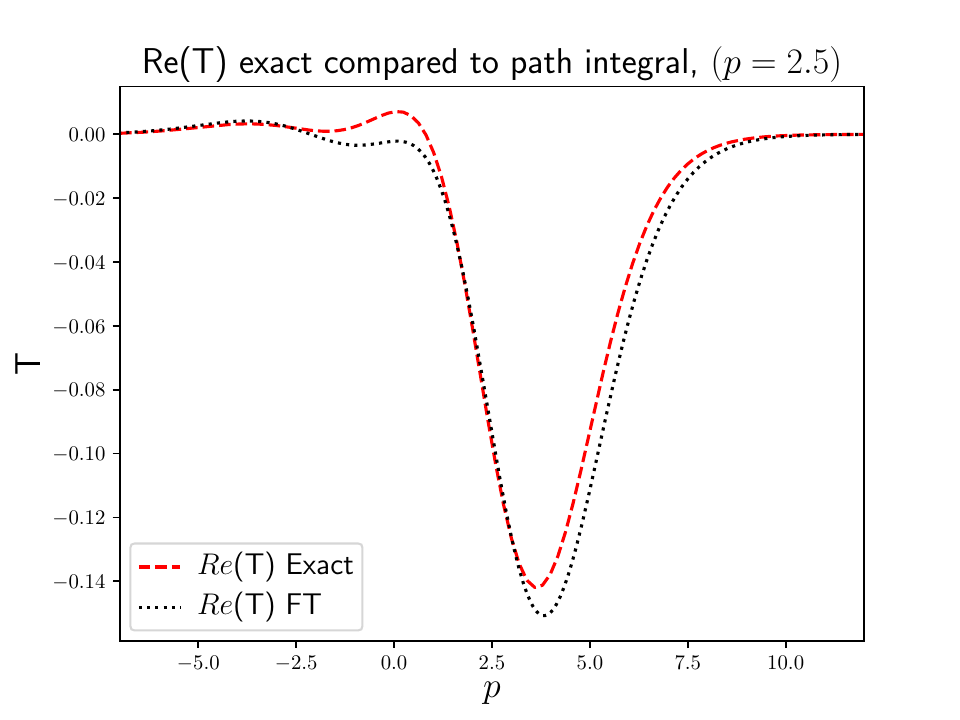}
\end{minipage}
\begin{minipage}[t]{.45\linewidth}  
\centering
\includegraphics[scale=.5]{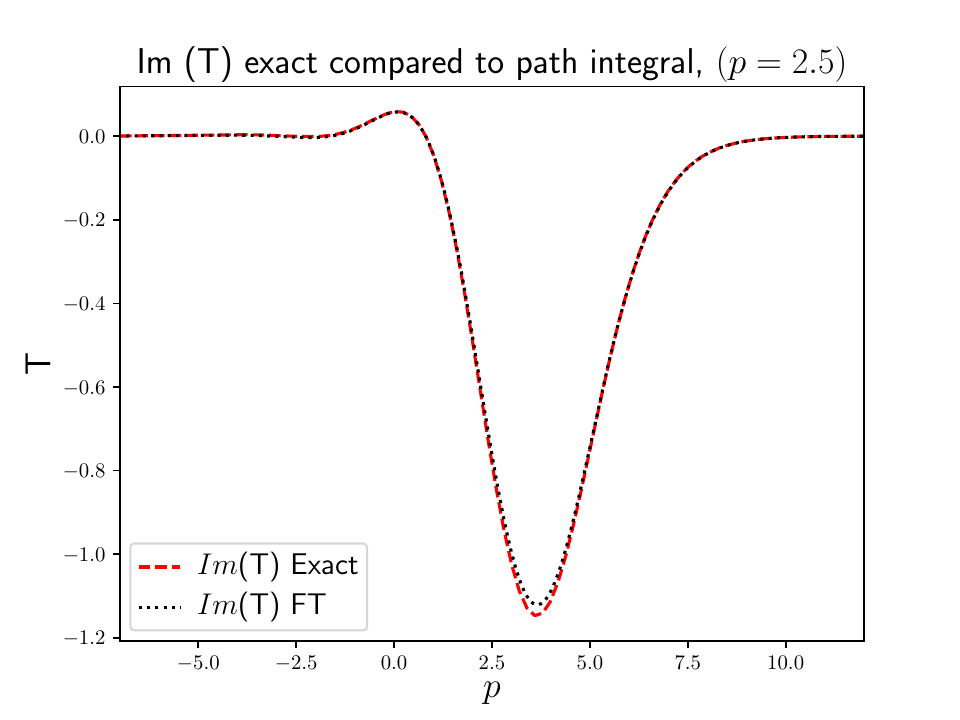}
\end{minipage}
\label{figure 18}
\end{figure}

The corresponding relative error in the real and imaginary
parts of the transition matrix at $p=2.5$ are shown in figures 19.
They are about $.03$ at the on shell point.

%
\begin{figure}
\caption{Plot of the absolute value of the relative error
of the real and imaginary parts of the half-shell transition matrix
path integral calculated using path integrals and
the Lippmann-Schwinger equation where $p=5.0$ is the on-shell point.}
\begin{minipage}[t]{.45\linewidth}  
\centering
\includegraphics[scale=.5]{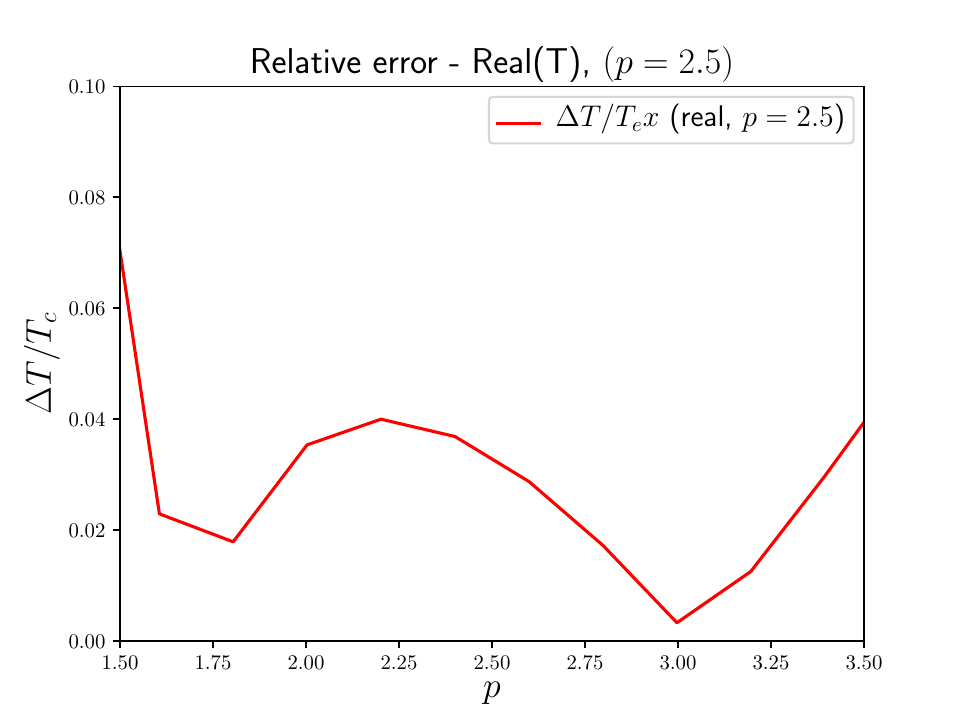}
\end{minipage}
\begin{minipage}[t]{.45\linewidth}  
\centering
\includegraphics[scale=.5]{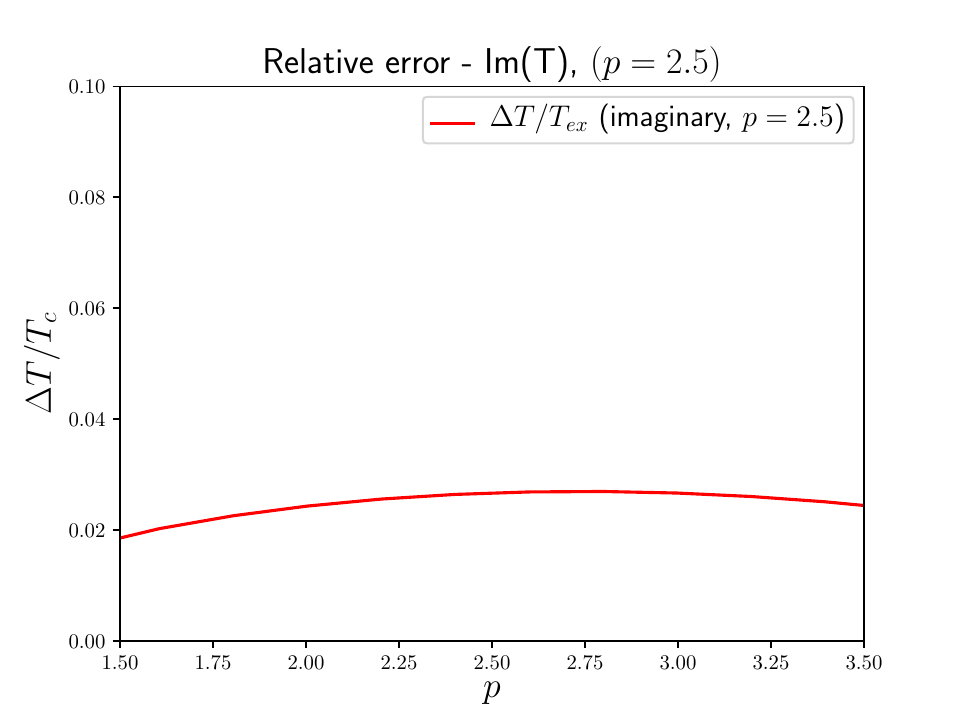}
\end{minipage}
\label{figure 19}
\end{figure}




The results indicate that lower energy calculations require more
computational resources than the higher energy calculations. There are
a number of reasons for this.  The slow moving wave packet must outrun
the negative spreading of the initial wave packet.  In addition, the
higher momentum resolution needed for low-energy calculations
increases the relevant volume that needs to be decomposed into
products of intervals.  The exhibited calculations, especially the
lower energy calculations, already involved iterating large matrices
($15000 \times 15000$). Fortunately these matrices do not have to be
stored.  The calculations are approximations in the sense that they
mathematically converge to the exact result.  This means that in
principle they can be improved by using more computational resources
(larger matrices and more time steps).  We did not attempt to go
beyond calculations of this size for $p=1.0$ and $p=2.5$ in part
because with the brute force computational methods used in this work,
small increases in resolution result in much larger calculations.
For this reason convergence is slow and it makes sense to first
investigate methods to improve the efficiency of the calculations,
rather than to continue investigating the limits of these inefficient
methods.  It is clear from the above discussions, that calculations of
zero-energy observables, like scattering lengths, would be very difficult. 
This is because it would be necessary to use narrow wave packets
in momentum that would result require large volumes.

\section{Summary and Conclusions}
 
This work demonstrated both the feasibility and challenges
of performing scattering calculations using real-time path integrals.

The exhibited calculations are based on a reinterpretation of the
path integral as the expectation of a potential functional with
respect to a complex probability distribution on a space of paths.
The space of paths is constructed by dividing the total time into a
large number of small time slices.  At each time slice space is
divided up into a large number of small windows, including two
semi-infinite windows.  The space of paths can then be decomposed into
disjoint equivalence classes where equivalent paths pass thorough the
same intervals at each time slice.
 
A complex probability is assigned to each equivalence class.  It is
constructed by decomposing the propagation of a free quantum
mechanical system into a sum of parts associated with each equivalence
class. Interactions are introduced by considering the effects of the
potential at each window.  Mathematically the time evolution of an
initial wave packet is represented by an average of a path-dependent
potential functional on the space of paths with respect to the complex
probability distribution.  The virtue of this method is that
it is a real approximation in the sense that it converges to
a global solution of the Schr\"odinger equation.

For the convergence of the calculations the equivalence class of paths
must be sufficiently fine that the potential locally constant
on each spatial volume element.    In addition, the time
steps must be sufficiently small for the Trotter product formula to
converge.  The fact that the complex probabilities are largely
supported on continuous paths means that for smooth short-range
potentials the relevant volume is finite. 

One difficulty is the large number of equivalence classes.  Each
equivalence class can be represented as the product of $N$ 
one-step classes for each time step.  This can be used to make an
approximate factorization of the $N$-time step complex probability as
a product of $N$ one-time-step probabilities.  This approximation
replaces the sum over the large number of equivalence classes by
computing powers of a large matrix, which was used in the
calculations.  In this representation the contribution of
each equivalence class is approximated by a particular sequence of
products of matrix elements.  For example, if $M_{mn}$ represents
the one step probability matrix, the product of numbers
\beq
M_{n_0 n_1} M_{n_1 n_2} \cdots M_{n_{19} n_{20}} 
\label{con:1}
\eeq
is approximately the complex probability for passing through the
sequence of windows $I_{n_{20}}$ at time $t_1$, $I_{n_{19}}$ at time
$t_2, \cdots $.  This factorization represents a tremendous increase
in efficiency - by using matrix algebra to treat a large number of
equivalence classes of paths in parallel.  Computationally
this looks like replacing the
path integral by successive applications of a transfer matrix to an
initial state, except in this case it is possible to identify and
extract the contribution of each cylinder set of paths to the
free-particle time-evolution operator. 

The scattering calculations were performed by approximating M{\o}ller
wave operators \cite{Moller} applied to normalizable wave packets.
This has the advantage of removing wave packet spreading effects from
the scattering calculations.  The relevant volume is the region where
the particles interact.  It is determined by the range of the
interactions and size of the free wave packets.  One possible strategy
to improve the efficiency would be to reduce the size of the active
volume by building some of the asymptotic properties of scattering
wave function into the free wave functions. This is essential for
long-range potentials \cite{Dollard} \cite{Mulherin}\cite{Busalev},
but similar methods could be applied to short range potentials.
 
The $p=5.0$ calculations presented in this paper formally involve
applying the $30^{th}$ power of a $5000 \times 5000$ matrix to a fixed
vector (for the transition matrix element calculations the $100^{th}$
power of a $10000 \times 10000$ matrix). This corresponds to averaging
over $5000^{30}$ (resp $(10000)^{100}$) equivalence classes of paths.
Because the one-step probabilities could be computed analytically,
accurately and efficiently, matrix elements could be computed on the
fly, which means that the computer storage required for these
calculations amounted to storing few complex vectors of length
5000(10000).  One of the surprising aspects of these calculations is
the stability of the sums over the complex probabilities.

The $p=5.0$ calculations presented were performed in minutes on a
desktop computer.  The $p=1.0$ and $p=2.5$ calculations required
larger space-time volumes and more time steps and took more than a day
on the same computer.
In terms of the number of
cylinder sets the $p=1.0$ calculations presented used the equivalent of
$(15000)^{1000}$ cylinder sets while the $p=2.5$ calculations used the
equivalent of $(16000)^{1000}$ cylinder sets.  As a result they took
considerably longer.  No attempt was made to be efficient.  All
intervals and time steps were taken to be the same size.  This is the
analog of computing a Riemann integral with equally spaced intervals.
There is a great deal of freedom both in how to choose intervals and
time slices that was not exploited.  Some  strategies that could be
explored would be use more resolution of cylinder sets near classical
paths.  The structure of the potential could be used to determine 
and optimal decomposition of each time slice into intervals.
The presented calculations simply applied the same matrix $N$ times to
the initial vector.  The application of the one-step probability
matrix to a localized vector could be made more efficient by
discarding small components of the resulting vector, reducing the size
of the vector that must be stored for each time step.  The one step
complex probabilities approximate a unitary transfer matrix for free
evolution, however exact unitarity is not forced on the resulting matrices.
Designing the one step probability so the matrix is exactly unitary
may reduce errors.

Beyond the numerical considerations, this framework is appealing in
that the input is a potential functional $F[\gamma]$; this picture 
is retained both exactly and in approximation.   This is in contrast to the 
usual path integral where the relevant weight functional formally 
looks like an action, but the terms that represent the time derivatives
have no legitimate interpretation as derivatives in the Trotter
product formula.  In the MNJ formulation, these terms do not appear 
explicitly; they are contained in the expression for the 
one-step probabilities

While the calculations in this paper are motivated by the 
complex probability interpretation, the computational strategy 
can be understood directly from Feynman's work.  His path integral 
results in the kernel $K(x,t;x',t')$ of the
time evolution operator (see equation (4.2) of \cite{Feynman_2})
\beq
\langle x \vert \psi (t) \rangle = \int K(x,t;x',t') dx' 
\langle x' \vert \psi (t') \rangle
\label{con:2}
\eeq
This can be expressed as the product of propagation over many 
time steps (see equation
(2.33) of \cite{Feynman_2}) :
\beq
\langle x \vert \psi (t) \rangle = \int K(x,t;x_1,t_1) dx_1
K(x_1,t_1;x_2,t_2) dx_2 \cdots K(x_{N-1},t_{N-1};x_N,t_N) dx_N 
\langle x_N \vert \psi (t_N) \rangle
\label{con:3}
\eeq
If the time intervals $t_{j+1}-t_j$, are sufficiently small then   
\beq
K(x_{j+1},t_{j+1};x_j,t_j) \approx
K_0(x_{j+1},t_{j+1};x_j,t_j)e^{-i V(x_j)(t_{j+1}-t_j)} 
\label{con:4}
\eeq 
where $K_0(x_{j+1},t_{j+1};x_j,t_j)$ is the free time-evolution 
kernel.  This is justified by the Trotter product formula.
Finally, if the integrals were replaced by numerical quadratures,
this would become
\[
\langle x \vert \psi (t) \rangle \approx \sum 
K_0(x,t ;x_{1n_1} ,t_1)e^{-i V(x_{1n_1})(t-t_1)} \Delta_{x_{1n_1}}
K_0(x_{1n_1},t_1 ;x_{2n_2} ,t_2)e^{-i V(x_{2n_2})(t_1-t_2)} \Delta_{x_{2n_2}}
\cdots \times
\]
\beq
K_0(x_{(N-1)n_{N-1}},t_{N-1} ;x_{Nn_N} ,t_N)e^{-i V(x_{Nn_N})(t_{N-1} -t_N)} \Delta_{x_{Nn_N}} \langle x_N \vert \psi (t_N) \rangle
\Delta_{x_{2n_2}}.
\label{con:5}
\eeq
The approximation
\beq
K_0(x_{1n_1},t_1 ;x_{2n_2} ,t_2)\Delta_{x_{2n_2}}
\approx 
\int_{x_{2n_2}-\Delta_{x_{2n_2}}/2}^{x_{2n_2}+\Delta_{x_{2n_2}}/2}
K_0(x_{1n_1},t_1 ;x ,t_2)dx
\label{con:6}
\eeq
corresponds to  the one-step probability  used in this work.   If this
replacement  is made  in  (\ref{con:5}) the  result  is equivalent  to
(\ref{njm.8}).  The important features to emphasize are (1) because of
the small time steps, the potential can be factored out and evaluated
at one of the quadrature points  and (2) the free  particle kernel is
known.  It is these two features that make  non-trivial calculations
possible.

The short time kernel
\[
K_0(x,t;x',t+\Delta t) 
\]
can be computed analytically leading to the following
unitary kernel for the approximate transfer matrix:
\[ 
TM(x,t;x',t+\Delta t) = 
K_0(x,t;x',t+\Delta t)e^{-i V(x')}
\]
\[
P(x'\to x, dx, \Delta t) := dx TM(x,t;x',t+\Delta t) 
\]
can be interpreted as a one-step complex probability
for a particle initially at $x'$ to end up within
$dx$ of $x$ in time $\Delta t$.  A direct numerical calculation
with this kernel may be more efficient that the brute force
calculation illustrated above.

The interesting question is how this method scales to more
particles or fields?  While
it is difficult to answer this question
without performing a test calculation and taking advantage of efficiencies,
For small time steps the Trotter formula justifies the replacements
\[
e^{-i (\sum H_i + \sum V_{ij})\Delta t} \to
\prod e^{-i H_i \Delta t} e^{-i \sum V_{ij} \Delta t} \to
\prod e^{-i H_i \Delta t} \prod e^{-iV_{ij} \Delta t} .
\]
In this form there are more products for a given time step, but the
non-trivial products only involve exponentials of single pairwise
potentials or one-body free propagators. Each of these operators,
$e^{-i H_i \Delta t}$ or $e^{-iV_{ij} \Delta t}$, only
act on the variables associated with particle $i$ or particles $i$ and $j$ in
the many-body wave function.  This suggest that the non-trivial part a
few-body calculation scales with the number of interacting pairs.
There are other factors, in that both the relevant volume and
interaction time will increase as more particles are added.  In
addition serious challenges with storing the vector
representing the evolved many-body wave function must be overcome..

The exhibited one-dimensional calculations are of comparable
difficulty to computing three-dimensional two-body scattering
calculations using partial waves in the center of mass frame.
Time-dependent methods based on numerical computation of wave
operators have been successfully applied to treat-three body $n$-$D$
breakup with strong $+$ Coulomb interactions \cite{kroger}.  These
calculation suggest that with some improvements in efficiency that
real path integral scattering calculations involving two or three
particles might be possible.

While this work demonstrates that scattering observables can be
directly computed using real-time path integrals, the calculations
used in this work cannot compete with conventional computational
methods.  These methods may provide a useful laboratory
for investigating the efficiency of different approximation
methods for treating real time evolution, which is needed
for scattering calculations on a quantum computer.

Conceptually, the one-step probability multiplied by the one-step
potential functional, $e^{-i V(y_i)\Delta t}$, which is the key to the
computational method, is essentially a transfer matrix, which is a
unitarized version of using the Hamiltonian to solve the Schr\"odinger
equation for finite time by taking many small time steps.  This has a
lot in common with evolving a product of smeared field operators with
the Heisenberg equations of motion.  Also, the one-step complex
probability is essentially free propagation over short time, which is
also well understood in the field-theory case.

One of the authors (WP) would like to acknowledge the generous
support of the US Department of Energy, Office of Science, 
grant number DE-SC0016457 who supported this research effort.

\bibliography{henstock_paper.bib}
\end{document}